\documentclass{jpp}

\usepackage{graphicx}
\usepackage{epstopdf}
\usepackage{xcolor}
\usepackage{amsmath}
\usepackage{hyperref}
\usepackage[export]{adjustbox}
\usepackage{float}

\usepackage{xspace}

\usepackage{comment}

\newcommand{\vect}[1]{\boldsymbol{#1}}

\usepackage{unicode}
\newcommand{\csso}{\fontencoding{LECO}\selectfont\char215}
\newcommand{\iotaslash}{\hspace*{0.1em}\iota\hspace*{-0.45em}\text{\csso}}

\title{Direct construction of stellarator-symmetric quasi-isodynamic magnetic configurations}
\author{Katia Camacho Mata \aff{1}\corresp{\email{katia.camacho@ipp.mpg.de}}, Gabriel G. Plunk\aff{1} and Rogerio Jorge\aff{1}
}
\affiliation{\aff{1}Max-Planck-Institut f\"ur Plasmaphysik, EURATOM Association, 17491 Greifswald, Germany
}

\begin{document}
\maketitle

We develop the formalism of the first order near-axis expansion of the MHD equilibrium equations described in \cite{garren1991magnetic}, \cite{plunk2019direct} and \cite{plunk2021direct}, for the case of a quasi-isodynamic, N-field period, stellarator symmetric, single-well magnetic field equilibrium. The importance of the magnetic axis shape is investigated, and we conclude that control of the curvature and torsion is crucial to obtain omnigenous configurations with finite aspect ratio and low effective ripple, especially for a higher number of field periods. For this reason a method is derived to construct classes of axis shapes with favourable curvature and torsion. Solutions are presented, including a three-field-period configuration constructed at an aspect ratio of $A{=}20$, with a maximum elongation of $e{=}3.2$ and an effective ripple under $1\%$, which demonstrates that high elongation is not a necessary feature of QI stellarators. 

\section{Introduction}
 
 In recent years, the W7-X experiment has demonstrated that intricately optimised stellarators can be successfully built and operated \citep{pedersen2018first,beidler2021demonstration}. Stellarators have attractive qualities, such as little net toroidal current and capacity for steady-state operation, that make them suitable for reactors. But, unlike in tokamaks, confinement  is not inherently good due to the three-dimensional geometry of the magnetic field. Stellarator magnetic fields need to be designed carefully to ensure neoclassical transport is sufficiently low. A sufficient condition for good orbit confinement in a stellarator is that of omnigenity,
 \begin{equation}\label{eq:omnigenity}
     \int ({\bf v}_{d}\cdot \nabla \psi) \ dt  = 0,
 \end{equation}
 where ${\bf v}_{d}$ is the total drift, and the integration is done over the bounce time of a trapped particle. Magnetohydrodynamic (MHD) equilibria for which trapped particle motion fulfils Eq. (\ref{eq:omnigenity})  have collisionless orbits that are radially confined. A subset of omnigenous magnetic fields are those that satisfy quasi-symmetry, in which the strength of the magnetic field is symmetric (axially, helically or poloidally) in magnetic coordinates. An example of omnigenous fields that are not quasi-symmetric are those called quasi-isodynamic (QI), which have poloidally closed contours of the magnetic field strength, as well as vanishing bootstrap currents at low collisionality \citep{,Helander_2009,helander2014theory}. The present work concerns this class of stellarators. 
 
Identification of good configurations, e.g. those with low neoclassical transport and easily buildable coils, is traditionally done through a two-step optimisation procedure. In the first step, a plasma boundary is deformed and  the physical properties of the resulting equilibrium are assessed using various codes until the desired plasma properties have been found. Then, in a second optimisation process, coils that reproduce the desired plasma boundary are sought.\footnote{The MHD equilibrium of a toroidal plasma with simply nested flux surfaces is determined by the shape of the boundary and the pressure and current profiles \citep{KK,helander2014theory}.} The equilibrium optimisation procedure requires solving computationally expensive codes in each step and prescribing an initial plasma boundary \citep{Henneberg-2021-a}. Such optimisation procedures generally identify local minima \citep{Henneberg-2021}, and it is therefore possible that other, lower,  minima may be found elsewhere in the parameter space. Additionally, QI optimisation often results in highly shaped plasma boundaries. In order to generate such plasma shapes, complicated coils may be necessary, which are difficult and expensive to build. It is not clear whether this complexity is inherent to particular QI equilibria or if it may be overcome by a more exhaustive or differently initialized search of parameter space. The near-axis approach, pursued here, has the potential to do just this.  
 
 This method, first introduced by \cite{garren1991magnetic} and revitalized by \cite{landreman2018direct} and \cite{landreman2019direct}, allows the construction of omnigenous solutions of the MHD equations at first order in the distance from the axis, using Boozer coordinates. Starting with an axis shape and a set of functions of the toroidal angle, this near-axis expansion (NAE) allows the systematic and efficient construction of plasma boundaries corresponding to omnigenous equilibria. Different procedures apply for the case of QI and for quasi-symmetry, with the latter described in \cite{landreman2019direct} and used successfully combined with an optimisation procedure in \cite{giuliani2022single}. The case of quasi-isodynamicity was first discussed by \cite{plunk2019direct} and will be described in greater detail in section \ref{Axis Expansion}.  

In section \ref{NfieldPeriods_symmetry} we derive all the constraints on the input functions for the case of a stellarator-symmetric, single-magnetic-well\footnote{By a "magnetic well" we do not refer to the concept from MHD stability theory, but, instead, to a trapping domain defined by the strength of the magnetic field along. Thus, a "single magnetic well" implies one maximum and one minimum in $B$ per field period.} solution with multiple field periods. We then proceed to find expressions for the geometric functions required as input for the near-axis construction and show that, for this particular case, the number of free parameters is reduced. 

Exactly omnigenous fields are necessarily non-analytic as shown by \cite{cary1997omnigenity}. Hence, omnigenity must necessarily be broken to achieve analytical solutions. In the near-axis expansion we do this in a controlled way, through the careful definition of one of the geometrical input functions, $\alpha(\varphi)$. A new way to express this function, more smoothly than done previously by \cite{plunk2019direct}, is proposed in section \ref{New_alpha}.

Very recently, single-field-period QI configurations with excellent confinement properties have been found by optimising within the space of these NAE solutions in \cite{Jorge2022}. Configurations with more field periods have proved challenging to find, even when using the optimiser. 

To investigate the causes of this difficulty, we analyse possible choices of the magnetic axis shape. Contrary to traditional optimisation, where a plasma boundary is the starting point and a magnetic axis has to be found as part of the equilibrium calculation, the near-axis expansion requires the magnetic axis shape as input. In section \ref{Axis_Shape} we discuss the freedom in the shape of the axis and its mathematical description. We shed some light on the difficulties associated with finding closed curves compatible with the NAE, specifically the need of finding closed curves with points of zero curvature at prescribed locations, as well as the role curvature and torsion play on the quality of the approximation. We then describe a method for generating stellarator-symmetric axis shapes with points of zero curvature and torsion at prescribed orders. 

A common problem encountered when constructing NAE solutions is the tendency that the boundary cross-section becomes highly elongated. Indications that increasing elongation of the plasma boundary leads to increasingly complex coils have been found by \cite{hudson2018differentiating}. Another argument for reducing elongation is the fact that very elongated cross-sections result in small plasma volume, thus increasing transport losses and construction costs per unit plasma volume. The relation between elongation and the relevant near-axis expansion parameters is  discussed in section \ref{Axis_Shape}, in particular the effect of torsion on shaping. 

In section \ref{example_2FP} we show how the NAE method can be used to construct a two-field-period solution that closely approximates omnigenity, and we discuss its geometric properties and neoclassical transport as measured by the effective ripple $\epsilon_{\mathrm{eff}}$. The particular case of a family of solutions with axes of constant torsion and increasing number of field periods is analysed in \ref{Constant_Torsion_QI} to show the impact of torsion on the quality of the near-axis approximation. Finally, in section \ref{3FP_example}, a three-field-period equilibrium is constructed around an axis shape specially designed to have low torsion in order to illustrate the importance of the axis shape in controlling transport and elongation.  
 
\section{Near-Axis Expansion}\label{Axis Expansion}

The near-axis expansion, as described by \cite{garren1991magnetic}, solves the equilibrium MHD equations by performing a first-order Taylor-Fourier expansion about the magnetic axis using Boozer coordinates $(\psi,\theta,\varphi)$. The general form of the magnetic field at first order in the distance from the axis is given by 
\begin{equation}\label{eq:magneticField}
    B(\epsilon, \theta, \varphi) \approx B_{0}(\varphi) \left(1 + \epsilon d(\varphi) \cos[\theta - \alpha(\varphi) ] \right),
\end{equation}
with the expansion parameter 
\begin{equation}
    \epsilon=\sqrt{\psi}\ll 1,
\end{equation} being a measure of the distance from the axis. As a consequence, these solutions will only be valid in the large aspect ratio limit. 

When imposing the conditions for omnigenity in the near axis expansion, as done by \cite{plunk2019direct}, the problem of constructing quasi-isodynamic magnetic equilibria can be transformed into the problem of specifying an on-axis magnetic field strength $B_{0}$, an axis shape, two functions $\alpha(\varphi)$ and $d(\varphi)$, which will be discussed in more detail later in this work, and solving a differential equation for $\sigma$ 
\begin{equation}\label{eq:sigma}
    \sigma^{'} + (\iotaslash -\alpha^{'})\left(\sigma^{2}+ 1 +\frac{B_{0}^{2}{\Bar{d}}^{4}}{4} \right)-G_{0}\Bar{d}^{2}\left(\tau + I_{2}/2\right) = 0,
\end{equation}
with $\iotaslash$ the rotational transform,  $G_{0}$ and  $I_2$ related to the poloidal and toroidal current, respectively, and $\Bar{d}(\varphi) = d(\varphi) /\kappa^{s}(\varphi)$. Primes represent derivatives with respect to the toroidal angular coordinate $\varphi$. In the case of stellarator symmetry, $\sigma(0) = 0$.

The magnetic axis is a space curve characterized by its torsion $\tau$ and signed curvature $\kappa^{s}$. The signed Frenet-Serret frame, $(\mathbf{t}, \mathbf{n}^{s}, \mathbf{b}^{s})$, of this curve is used as an orthonormal basis to describe the coordinate mapping. The tangent vector $\mathbf{t}$ is identical to that of the traditional Frenet-Serret apparatus, and the normal and binormal signed vectors, $\mathbf{n}^{s}, \mathbf{b}^{s}$, as well as the signed curvature change signs at points where $\kappa = 0$. Within this framework the position vector to first order is given by
\begin{equation}
    \mathbf{x} \approx \mathbf{r}_{0} + \epsilon \left( X_{1}\mathbf{n}^{s} + Y_{1}\mathbf{t}^{s}, \right)
\end{equation}
where $\mathbf{r}_{0}$ describes the position of the magnetic axis.

By specifying a distance from the axis, and after finding $\sigma$, an approximate plasma boundary can be described by 
\begin{gather}\label{eq:X0_Y0}
    X_{1} = \frac{d(\varphi)}{\kappa^{s}} \cos{[\theta - \alpha (\varphi) ]} \\
    Y_{1} = \frac{2\kappa^{s}}{B_{0}d(\varphi)} \left( \sin{[\theta - \alpha (\varphi)]} + \sigma(\varphi) \cos{[\theta - \alpha (\varphi) ]}   \right). \label{eq:Y1}
\end{gather}

The functions $\alpha(\varphi)$ and $d(\varphi)$ need to be prescribed for the construction and are required to satisfy specific conditions to correspond to omnigenous solutions. 
In order to formulate these conditions mathematically, a certain coordinate mapping described in  \cite{cary1997omnigenity} was used in \cite{plunk2019direct}. For each magnetic well in the on-axis magnetic field, trapping domains are delimited by the maxima defining the well. This is divided into a right-hand domain $D_{R}$ and a left-hand domain $D_{L}$, depending on which side of the minimum of the well the points lie. For each point $\varphi$ in a left-hand domain we identify its corresponding bounce point $\varphi_{b}(\varphi)$ in the right-hand domain by the condition  
\begin{equation}
    B_{0}(\varphi) = B_{0}(\varphi_{b}(\varphi)),
\end{equation}
and vice versa. It is clear from this construction that
\begin{equation}
    \varphi_{b}(\varphi_{b}(\varphi)) = \varphi.
\end{equation}
The angular distance between an arbitrary point $\varphi$ and its corresponding bounce point is given by
\begin{equation}\label{eq:Delta_Phi}
    \Delta \varphi (\varphi) \equiv \varphi - \varphi_{b}(\varphi).
\end{equation}

\cite{plunk2019direct} provided a way to construct the functions $d(\varphi)$ and $\alpha(\varphi)$ in $D_{R}$, giving their dependency in $D_{L}$, to guarantee  omnigenity. Equating the near-axis and omnigenous forms of $B_{1}$ gives 
\begin{equation}\label{eq:6.9_QI_Paper}
    B_{0}(\varphi) d(\varphi) \cos{(\theta - \alpha(\varphi))} =- F_{1}(\theta,\varphi) B_{0}^{\prime}(\varphi), 
\end{equation}
where $F_{1}(\theta,\varphi)$ satisfies the following symmetry
\begin{equation}\label{eq:F_omnigenous_6.10}
    F_{1}(\theta,\varphi) = F_{1}(\theta,-\iotaslash \Delta\varphi(\varphi)),\quad \text{for   } \varphi \in D_{R}.
\end{equation}
Solving equation (\ref{eq:6.9_QI_Paper}) for $F_{1}$ and plugging it in equation (\ref{eq:F_omnigenous_6.10}) we obtain the following set of conditions 
\begin{equation}\label{eq:d}
    d(\varphi) = \varphi^{'}_{b}(\varphi) d(\varphi_{b}(\varphi)),\quad\text{for }\varphi\in  D_{R}
\end{equation}
\begin{equation}\label{eq:alpha}
    \alpha(\varphi) = \alpha(\varphi_{b}(\varphi)) + \iotaslash \Delta \varphi(\varphi),\quad\text{for }\varphi\in  D_{R}.
\end{equation}

Additionally, $d$ must vanish at all extrema of the on-axis magnetic field $B_{0}$, as seen from equation (\ref{eq:6.9_QI_Paper}). The curvature of the magnetic axis $\kappa^{s}$ needs to have zeros of the same order as $d$ at these points for the plasma boundary to be well described, i.e. so that $X_{1}$ and $Y_{1}$, which are proportional to $d/\kappa^{s}$ and $\kappa^{s}/d$, respectively, remain non-zero and bounded.  

It is important to notice that periodicity cannot be enforced if the condition (\ref{eq:alpha}) on $\alpha(\varphi)$ is satisfied and the rotational transform $\iotaslash$ is irrational. We can see this from evaluating (\ref{eq:alpha})  at the maximum $\varphi = 2\pi$,
\begin{equation}\label{eq:alpha_periodicity}
    \alpha(2\pi) -\alpha(0) = 2\pi \iotaslash.
\end{equation}
However, continuity of $B_{1}$, $X_{1}$ and $Y_{1}$ requires 
\begin{equation}\label{eq:alpha_continuityBX}
    \alpha(2\pi) - \alpha(0) = 2\pi M,
\end{equation}
with $M$ defined as the number of times the axis curvature vector $\mathbf{n}^{s}$ rotates during one toroidal transit. Thus, equation (\ref{eq:alpha_periodicity}) is generally in conflict with Eq.~(\ref{eq:alpha_continuityBX}) and omnigenity is only consistent with continuity for integer values of $\iotaslash$. In \cite{plunk2019direct} this conflict was resolved by introducing small matching regions around $\varphi=0$ and  $\varphi = 2\pi$, where omnigenity is abandoned and $\alpha$ is defined to guarantee periodicity. A different approach to solving this problem, as well as the form these conditions take for a single well, $N$-field-period, stellarator-symmetric configuration will be discussed in the following sections.   

\section{$N$-field periods and stellarator symmetry}\label{NfieldPeriods_symmetry} 
A magnetic field is said to be stellarator-symmetric if it is invariant under a 180-degree rotation (the operations $I_n$ defined below) around some axis perpendicular to the vertical axis. This type of symmetry was defined formally by \cite{dewar1998stellarator}, who introduced a symmetry operator $I_{0}$ by
\begin{equation}
  I_{0}\mathbf{F}(\theta,\varphi) =  \mathbf{F}(-\theta,-\varphi),
\end{equation}
with $\theta$ and $\varphi$ being angular coordinates. We say that a vector $\mathbf{F}$ possesses stellarator symmetry if
 \begin{equation}
  I_{0}[F_{\psi},F_{\theta},F_{\varphi}] = [-F_{\psi},F_{\theta},F_{\varphi}], 
\end{equation}
 where $F_{j}=\frac{\partial{\mathbf{x}}}{\partial j} \cdot \mathbf{F}$, with $j=\{\psi,\theta,\varphi\}$, are the covariant components of $\mathbf{F}$. For the case of a scalar quantity, for instance $|F|$, stellarator symmetry implies
 \begin{equation}
     I_{0}|F| = |F|.
 \end{equation}
If the vector field $\mathbf{F}$ possesses $N$-fold discrete symmetry about the $z$-axis, then stellarator symmetry also exists about the cylindrical inversion symmetry operation with respect to the half-line $\{\phi = 2i\pi/N, Z= 0\}$, with $i= 1,2,...,(N-1)$, and with respect to the half-line in the middle of each field period $\{\phi = (2i-1)\pi/N, Z= 0\}$, with $i= 1,2,...,N$. These symmetry operators will be referred to as $I_{2i}$ and $I_{2i-1}$, respectively. 

\begin{figure}
    \centering
    \includegraphics[width=0.7\textwidth]{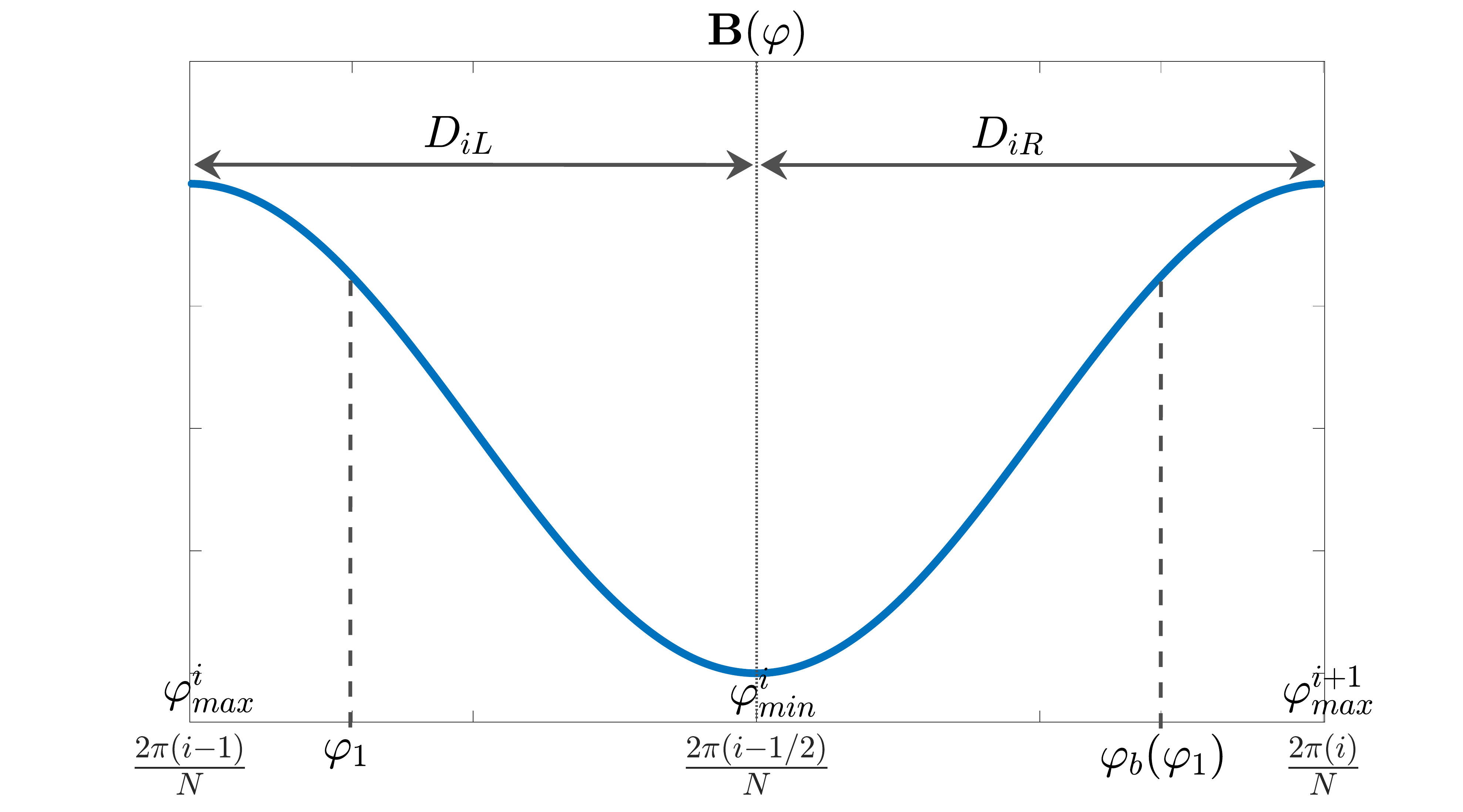}
    \caption{One well magnetic field example. The right and left hand domains, $D_{iR}$ and $D_{iL}$ are indicated. The angular position of the minimum and maxima of the well are labeled by $\varphi_{min}^{i}$ and $\varphi_{max}^{i}$, respectively.} 
    \label{fig:B_domains}
\end{figure}

Let us now consider the case of a stellarator-symmetric $N$-field-period quasi-isodynamic configuration, and focus on the case of one magnetic well per field period. In order to be stellarator symmetric, the magnetic field strength ${B}$ needs to fulfil 
\begin{equation}
    B(\theta,\varphi_{\text{min}}^{(i)}+ \delta \varphi) = B(-\theta,\varphi_{\text{min}}^{(i)}- \delta \varphi), 
\end{equation}
 which is obtained by invoking the symmetry operator $I_{2i-1}$. The angular position of the $i$-th minimum, $\varphi_{min}^{(i)}$ is located on the rotation axis, and is given by
 \begin{equation}
     \varphi_{\mathrm{min}}^{(i)} = \frac{(2i-1)\pi}{N},\quad\text{for }i = 1,...,N
 \end{equation}
while the $i$-th maximum is
 \begin{equation}
     \varphi_{\mathrm{max}}^{(i)} = \frac{2(i-1)\pi}{N},\quad\text{for }i = 1,...,N.
 \end{equation}
The trapping domain in each period is labelled with the subscript $i$, shown in figure \ref{fig:B_domains}, and the left- and right-hand domains are defined as 
\begin{align*}
    D_{iL}(\varphi), &\quad\text{for     }\tfrac{2\pi(i-1)}{N}\leq \varphi \leq \tfrac{(2i-1)\pi}{N} \\
    \\
    D_{iR}(\varphi), &\quad\text{for    } \tfrac{(2i-1)\pi}{N}\leq \varphi \leq  \tfrac{2\pi i}{N}   
\end{align*}
Now, equation (\ref{eq:Delta_Phi}), which gives the distance between bounce points, can be written as 
\begin{equation}\label{eq:deltaPhi_omn}
    \Delta\varphi(\varphi) = 2(\varphi - \varphi_{min}^{(i)})
\end{equation},
and the bounce points can be found using 
\begin{equation}\label{eq:BouncingPoints}
    \varphi_{b}(\varphi) = \varphi - \Delta \varphi(\varphi) = 2\varphi_{min}^{(i)} - \varphi.
\end{equation}
Using this new notation, we can write the symmetry operator $I_{2i-1}$ as 
\begin{equation}\label{eq:symmetryOperation}
   I_{\varphi_{min}^{i}} f(\psi,\theta,\varphi)= f(\psi,-\theta,\varphi_{b}).
\end{equation}

Let us now focus on the condition for omnigenity on the function $d(\varphi)$, which is shown in Eq.~(\ref{eq:d}). We substitute $\varphi_{b}$ from expression (\ref{eq:BouncingPoints}) and notice that $\varphi_{b}^{'}=-1$, to find 
\begin{equation}\label{eq:d_omni}
    d(\varphi) = -d(\varphi_{b})= - d(2\varphi_{min}^{(i)} - \varphi),\quad\text{for }\varphi\in  D_{iR}.
\end{equation}
As a consequence, $d(\varphi)$ must necessarily be an odd function with respect to $\varphi_{min}^{(i)}$, the bottom of the well. Note that $\varphi_{b}^{\prime}(\varphi)$ is generally negative near bounce points for $\varphi \in D_{iL}$, even without requiring stellarator symmetry, so the order of the zeros of $d$ must always be odd, and therefore the same is true for those of $\kappa^{s}$. 

For the case of $\alpha(\varphi)$, we first define a quantity $\Delta \alpha$ in the same fashion as Eq.~(\ref{eq:Delta_Phi})
\begin{equation}
    \Delta \alpha(\varphi) \equiv \alpha(\varphi) - \alpha(\varphi_{b}(\varphi)). 
\end{equation}
By replacing $\alpha$ by expression (\ref{eq:alpha}) in the previous definition, we obtain 
\begin{equation}
    \Delta \alpha(\varphi) = \iotaslash\Delta\varphi(\varphi), 
\end{equation}
and using expression (\ref{eq:deltaPhi_omn}) we get
\begin{equation}
     \Delta \alpha(\varphi) = 2\iotaslash (\varphi - \varphi_{min}^{(i)}).
\end{equation}
We see that $\alpha$ needs to have a part proportional to $\iotaslash \Delta \varphi(\varphi)$, to guarantee the right form of $\Delta\alpha$, and an even part $\alpha_{e}$ such that 
\begin{equation}
    \alpha_{e}(\varphi) = \alpha_{e}(\varphi_{b}(\varphi)). 
\end{equation}
Accordingly, we write $\alpha$ as 
\begin{equation}\label{eq:alphaForm}
    \alpha(\varphi) = \tfrac{1}{2}\iotaslash\Delta\varphi(\varphi) + \alpha_{e}(\varphi).
\end{equation}
In order to find $\alpha_{e}$, we notice that the first-order correction to the magnetic field, given in Eq.~(\ref{eq:magneticField}),
\begin{equation}\label{eq:B_firstOrder}
    B_{1} =  B_{0}(\varphi) d(\varphi)\cos{(\theta-\alpha(\varphi))}
\end{equation}
 needs to possess stellarator symmetry. We know from (\ref{eq:d_omni}) that $d(\varphi)$ is an odd function and, by construction, that $B_{0}$ is an even function. Therefore $\cos{(\theta-\alpha(\varphi))}$ must be odd under the symmetry operation defined in (\ref{eq:symmetryOperation}), requiring
\begin{equation}
    \cos{(\theta - \alpha(\varphi_{min}^{i}))}=0,
\end{equation}
which is valid for any integer $n_{i}$ such that
\begin{equation}
    \alpha(\varphi_{min}^{i}) = \pi (n_{i}+ \tfrac{1}{2}).
\end{equation}
Evaluating equation (\ref{eq:alphaForm}) at the minimum, we find 
\begin{equation}
    \alpha_{e} = \pi(n_{i}+ \tfrac{1}{2}),
\end{equation}
so that $\alpha$ takes the form 
\begin{equation}\label{eq:alpha_2}
    \alpha(\varphi) = \iotaslash (\varphi - \varphi_{min}^{(i)}) +  \pi(n_{i}+ \tfrac{1}{2}). 
\end{equation}
Now let us impose a necessary (but not sufficient) condition   for periodicity of $B$ from one field period to the next, i.e. 
\begin{equation}\label{eq:alpha_max}
    \alpha(\varphi_{max}^{i+1})-\alpha(\varphi_{max}^{i}) = 2\pi m,  
\end{equation}
where $m$ is the number of times the signed curvature vector $\textbf{n}^{s}$ rotates per field period. This ensures that the poloidal angle $\theta$ is periodic and increases by $2\pi$ after a full toroidal rotation. The addition of the term $2\pi m$ to $\alpha$ compensates the poloidal rotation of the axis (measured by m) since $\alpha$ effectively behaves as a phase-shift on $\theta$. The previous expression translates into a relation for $n_{i}$
\begin{equation}
    \pi(n_{i+1}+1/2)-\pi(n_{i}+1/2) = 2\pi m,
\end{equation}
which can also be written as
\begin{equation}
    n_{i} = 2im+n_{1}.
\end{equation}
Without loss of generality we can choose $n_{1}=0$, and inserting this result in Eq.~(\ref{eq:alpha_2}) we obtain an expression for $\alpha(\varphi)$ satisfying omnigenity and N-fold periodicity
\begin{equation}\label{eq:alpha_Final}
    \alpha(\varphi) = \iotaslash (\varphi-\varphi_{min}^{i})+\pi (2 i m + \tfrac{1}{2}).
\end{equation}

The shape of the boundary must also be made periodic, which can be achieved in the same way as done in \cite{plunk2019direct}, by introducing so-called buffer regions around the maxima of $B_{0}$. In these regions, $\alpha$ is not calculated using (\ref{eq:alpha_Final}), but is instead chosen to give a periodic plasma boundary. The function in the buffer region needs to be constructed carefully to make sure the function itself and its derivatives are continuous and smooth in order to avoid numerical difficulties in equilibrium solvers such as VMEC \citep{hirshman1983steepest}. 

We note that although we have some freedom in the choice of $n_{1}$ and hence in the value of $\alpha$, the term entering equation (\ref{eq:sigma}) and defining the equilibria is $\alpha^{'}(\varphi)$, which is independent of the choice of $n_{1}$. The only freedom left in the choice of $\alpha(\varphi)$ is on how omnigenity is broken to impose continuity of the solutions.   

\section{Smoother $\alpha(\varphi)$}\label{New_alpha}

\begin{figure}
    \centering
    \includegraphics[width=1.0\textwidth]{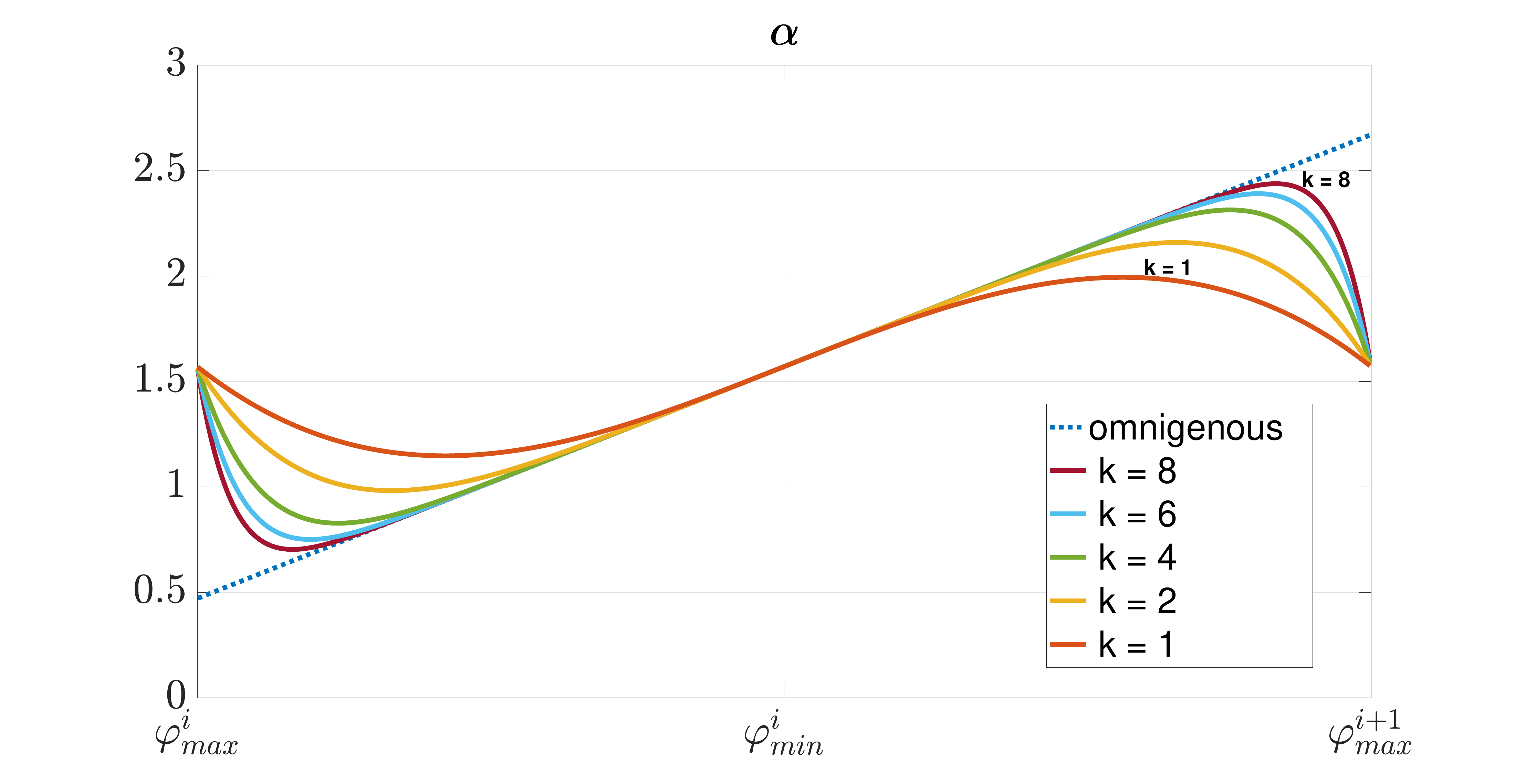}
    \caption{Behavior of the function $\alpha(\varphi)$ over one period, for the case $m=0$. The dotted line shows the fully omnigenous case, which is non continuous at $\varphi_{\mathrm{max}}$. Solid lines correspond to $\alpha$ given by Eq.~(\ref{eq:alpha_final}) with values for the parameter $k$ ranging from 1 to 8.}
    \label{fig:alpha}
\end{figure}

In order to avoid the problems derived from defining $\alpha(\varphi)$ as a piecewise function, we propose a different approach, namely by adding an omnigenity-breaking term to equation (\ref{eq:alpha_Final}) that allows periodic solutions 
\begin{equation}
    \alpha(\varphi) = \iotaslash (\varphi-\varphi_{min}^{i})+\pi (2 m i + \tfrac{1}{2}) + a(\varphi-\varphi_{min}^{i})^{2k+1}. 
\end{equation}
The last term in this expression goes to zero at the bottom of the well as long as $k\geq -1/2$ and hence makes the solution omnigenous for deeply trapped particles. The parameter $a$ can be chosen in such a way that periodicity of $\alpha(\varphi)$ is guaranteed. 
A first requirement is continuity at $\varphi_{max}^{(i)}$, i.e.
\begin{equation}
    \lim_{\varphi\to\varphi_{\mathrm{max}}^{(i+1)-}} \alpha(\varphi)  = \lim_{\varphi\to\varphi_{\mathrm{max}}^{(i+1)+}} \alpha(\varphi),
\end{equation}
where we are considering the $i$-th well when approaching from the left and the well labelled $i+1$ when approaching from the right. Hence we obtain
\begin{multline*}
     \iotaslash \left(\varphi_{\mathrm{max} }^{(i+1)}-\varphi_{ \mathrm{min}}^{(i)}\right) + a\left(\varphi_{\mathrm{max}}^{(i+1)}-\varphi_{\mathrm{min}}^{(i)}\right)^{2k+1} \\= \iotaslash \left(\varphi_{\mathrm{max}}^{(i+1)}-\varphi_{\mathrm{min}}^{(i+1)}\right) + a\left(\varphi_{\mathrm{max}}^{(i+1)}-\varphi_{\mathrm{min}}^{(i+1)}\right)^{2k+1} + 2\pi mi, 
\end{multline*}
and when inserting the values of $\varphi_{\mathrm{max} }^{(i)}$ and $\varphi_{\mathrm{min} }^{(i)}$
\begin{equation*}
    \iotaslash \left( - \pi / N\right) + a\left(-\pi / N\right)^{2k+1} + 2\pi m \\= \iotaslash \left(\pi / N\right) + a\left(\pi / N\right)^{2k+1},  
\end{equation*}
we find an expression for $a$
\begin{equation}
    a = \frac{\pi \left( m - \iotaslash /N \right)}{(\pi/N )^{2k+1}},
\end{equation}
which finally gives us an expression for $\alpha$ that breaks omnigenity in a smooth and controlled way 
\begin{equation}\label{eq:alpha_final}
    \alpha(\varphi) = \iotaslash (\varphi-\varphi_{\mathrm{min}}^{i})+\pi (2 m i + \tfrac{1}{2}) + \pi \left( m - \iotaslash /N \right) \left(\frac{\varphi-\varphi_{\mathrm{min} }^{i}}{\pi/N} \right)^{2k+1}. 
\end{equation}
In figure \ref{fig:alpha}, we can see the impact the choice of the parameter $k$ has over the shape of $\alpha(\varphi)$ for an axis shape with $m=0$. It is clear that increasing $k$ results in a function closer to that required for omnigenity but at the cost of a sharp behaviour close to $\varphi_{\mathrm{max}}$ to preserve periodicity. 

The equilibrium constructed using this $\alpha$ will be approximately omnigenous as long as the last term in equation (\ref{eq:alpha_Final}) is small relative to the first term, i.e.,

\begin{equation*}
\left|  \pi \left( m - \iotaslash /N \right) \left(\frac{\varphi-\varphi_{\mathrm{min} }^{i}}{\pi/N} \right)^{2k+1} \right| \ll \left|\iotaslash \left( \varphi - \varphi_{\mathrm{min}}^{(i)} \right)\right|,
\end{equation*}
which can be further simplified and rearranged as
\begin{equation}\label{eq:omn_condition}
    \left( \frac{\left| \varphi - \varphi_{\mathrm{min}}^{(i)}\right|}{\pi/N} \right)^{2k} \ll \frac{1}{\left| Nm/\iotaslash -1\right|}.
\end{equation}
 The left side of this expression includes $\left| \varphi - \varphi_{\mathrm{min}}^{(i)}\right|$, which is always smaller than $\pi /N$ in a well. 
 Hence the condition for omnigenity being achieved closely everywhere in the well, i.e. also for $\varphi -\varphi_{\mathrm{min}} \approx \pi /N$, reduces to 
\begin{equation}\label{eq:Nm_iota}
   ( Nm/\iotaslash -1) \ll 1.
\end{equation}

This implies $N m \sim \iotaslash$, which, as expected can only be satisfied for nearly rational $\iotaslash$. Although a rational value of $\iotaslash$ is inconsistent with confinement, it may be advantageous to seek nearly rational values as a strategy for finding almost omnigenous configurations. We also note that the $m=0$ case, in which (\ref{eq:Nm_iota}) is not technically satisfied, is however interesting since the limit of small $\iotaslash$ implies that the absolute size of $\alpha$ will remain small, as required for approximate omnigenity. 


A yet smoother choice of the function $\alpha(\varphi)$ with continuous derivatives up to third order at $\varphi_{\mathrm{max}}$ can be achieved by adding an extra term,
\begin{equation}
    \alpha_{{\mathrm{II}}}(\varphi) = \iotaslash (\varphi-\varphi_{\mathrm{min}}^{i})+\pi (2 m i + \tfrac{1}{2}) + a(\varphi-\varphi_{\mathrm{min}}^{i})^{2k+1} + b(\varphi-\varphi_{\mathrm{min}}^{i})^{2p+1}. 
\end{equation} 
For the configurations shown in this work we will use $\alpha(\varphi)$ as described in equation (\ref{eq:alpha_Final}), which appears sufficiently smooth for the cases we have studied, but more details about $\alpha_{\mathrm{II}}(\varphi)$ are discussed in Appendix I. 

\section{Axis Shape}\label{Axis_Shape}

The shape of the magnetic axis is perhaps the most important input for the construction of stellarator configurations since the axis properties seem to strongly affect the success of the construction, as measured by the accuracy of the approximation at finite aspect ratio. 

The need to have an axis with points of zero curvature has already been discussed, but additionally, low-curvature axes are attractive because they improve the accuracy of the near-axis approximation (indeed, the original work of Garren and Boozer defined the expansion parameter in terms of the maximum curvature of the magnetic axis), and are also associated with a low amplitude first-order magnetic field
\begin{equation*}
    B_1 = \epsilon B_0 \bar{d} \kappa^{s} \cos(\varphi - \alpha),
\end{equation*}
where the definition of $\bar{d}$ has been substituted in Eq.~(\ref{eq:B_firstOrder}).  Note that $\bar{d}$ cannot be made too small without causing large elongation, and therefore minimizing $\kappa^s$ is an effective strategy for improving omnigenity at finite $\epsilon$. 

It is also desirable to limit axis torsion, which enters directly into equation (\ref{eq:sigma}) for $\sigma$, as it will be clear in the following discussion. 
At first order, the cross-sections  of the plasma boundary at constant $\varphi$ are elliptical, with elongation $e$ defined as the ratio between the semi-major and semi-minor axes. \cite{landreman2018direct} derived an expression for elongation in terms of $X_{1}$ and $Y_{1}$ (eqn. B4). Using (\ref{eq:X0_Y0}) and (\ref{eq:Y1}), we obtain an expression for $e$  dependent on  the input parameters of the construction, namely $\bar{d}$, $B_{0}$, and $\sigma$,
\begin{equation}
     e = \frac{B_{0}\bar{d}^{2}}{4} + \frac{1}{B_{0}\bar{d}^2}(1+\sigma^2) + \sqrt{\left( \frac{B_{0}\bar{d}^{2}}{4}\right)^2 +  \frac{(\sigma^2-1)}{2} + \left(\frac{1}{B_{0}\bar{d}^2}(1+\sigma^2)\right)^{2}}.
\end{equation}
We can simplify the previous expression by introducing
\begin{equation}\label{eq:e_bar}
    \bar{e} = \frac{B_{0}\bar{d}^{2}}{2},
\end{equation}
leading to
\begin{equation}\label{eq:elongation}
     e = \frac{\bar{e}}{2} + \frac{1}{2\bar{e}}(1+\sigma^2) + \sqrt{ \frac{\bar{e}^{2}}{4} +  \frac{(\sigma^2-1)}{2} + \frac{(1+\sigma^2)}{4\bar{e}^2}}.
\end{equation}

One particular interesting limit is the idealized case of constant elongation. This can be achieved by choosing $\sigma$ constant. Given that, due to stellarator symmetry, $\sigma(0)=0$ then  $\sigma$ must be zero everywhere, which transforms Eq.~(\ref{eq:elongation}) into
\begin{equation}
    e = \frac{1}{2\bar{e}}\left( \bar{e}^{2} +1 + \left|\bar{e}^{2}-1\right| \right),
\end{equation}
this will result in a plasma boundary with constant elongation as long as $\bar{e}$ is independent of the toroidal angle $\varphi$. Finding plasma equilibria with constant elongation can be achieved  by choosing $d(\varphi)$ appropriately to ensure this condition is satisfied. 

Now, we can introduce the conditions that lead to constant elongation in equation (\ref{eq:sigma})
\begin{equation}\label{eq:sigma_constElong}
   (\iotaslash -\alpha^{'})\left( 1 +\bar{e} \right)=  \frac{2G_{0}\bar{e}}{B_{0}} \tau, 
\end{equation}
where we have specialized to the case of vanishing current density on axis, $I=0$, the standard situation for QI stellarators \citep{Helander_2009,Helander_2011}. Since omnigenity requires $\iotaslash - \alpha^\prime \approx 0$ (see Eq. \ref{eq:alpha_Final}), equation (\ref{eq:sigma_constElong}) implies that $\tau \approx 0$ is necessary for solutions with constant elongation. This shows that axes with low torsion are compatible with simple equilibrium boundary shapes. 

\subsection{Axis Construction}\label{sec:Axis_Construction}

A space curve's shape is entirely determined by its curvature $\kappa$ and torsion $\tau$. From these two quantities, it is possible to calculate the tangent, normal and binormal vectors $(\mathbf{t}, \mathbf{n}, \mathbf{b})$ using the Frenet-Serret formulas
\begin{eqnarray}\label{eq:Frenet-Serret formulas}
   \frac{d\mathbf{t}}{ds} =& \kappa(s)\mathbf{n}, \nonumber \\
   \frac{d\mathbf{n}}{ds} =& -\kappa(s)\mathbf{t} + \tau(s)\mathbf{b}, \\
    \frac{d\mathbf{b}}{ds} =& -\tau(s) \mathbf{n} \nonumber,
\end{eqnarray}
where $s$ is the arc length, used for parametrizing the curve. Then, numerical integration of the tangent vector, $\mathbf{t} = d\mathbf{r}/ds$, yields the curve described by $\mathbf{r}$. Unfortunately, prescribing periodic $\kappa$ and $\tau$ is not sufficient for finding closed curves. A parameter optimisation is thus needed to find curves that can be used as magnetic axes. This is the approach used to find the axes curves described in section \ref{Constant_Torsion_QI}.

Although a Frenet description seems optimal for controlling torsion and curvature precisely, a truncated Fourier series is advantageous for its simplicity, smoothness and the fact that such curves are automatically closed. Smoothness is especially important for obtaining solutions that remain accurate at lower aspect ratio.  Note that sharp derivatives $d/d\varphi \sim 1/\epsilon$ invalidate the near-axis expansion, which is a limit that is especially felt at high field period number. A method for generating simple axis curves, represented by a relatively small number of Fourier coefficients, is briefly outlined here and described in greater detail in Appendix II.

We represent the magnetic axis in cylindrical coordinates as 

\begin{equation}
    \mathbf{x} = \hat{\bf R}(\phi) R(\phi)  +  \hat{\bf z} z(\phi).
\end{equation}
The usual Fourier representation for a stellarator-symmetric axis is
\begin{eqnarray}
    R(\phi) = \sum_{n=0}^{n_\mathrm{max}} R_c(n) \cos(n N \phi),\label{R-Fourier}\\
    z(\phi) = \sum_{n=1}^{n_\mathrm{max}} z_s(n) \sin(n N \phi).\label{z-Fourier}
\end{eqnarray}
A local form is used to establish conditions on the derivatives of these functions about a point of stellarator symmetry (also coinciding with an extrema of $B_0(\phi)$), that can be then used to generate a linear system of equations for the Fourier coefficients.

Conditions on torsion and curvature, specifically zeros of different orders, are imposed locally and 
converted into constraints on the derivatives of the axis components $R$ and $z$, and then applied to a truncated Fourier representation. This results in a set of linear conditions on the Fourier coefficients that can be solved numerically, or by computer algebra. The orders of the zeros,
together with the set of Fourier coefficients, define a space that can be used for further
optimisation.

As a very simple example, one may consider a symmetric class of curves, as in \cite{plunk2019direct}
\begin{eqnarray}\label{eq:R_2param}
     R = 1 + R_c(2) \cos(2 N \phi),\\
     z = z_s(2) \sin(2 N \phi).\label{eq:z_2param}
\end{eqnarray}
Due to the fact that only the even mode numbers are retained ($z_s(1) = R_c(1) = 0$), the condition for zeros of curvature at first order need only be applied at $\phi = 0$, {\em i.e.} $R^{\prime\prime}(0) = R(0)$, and the result is

\begin{equation}\label{eq:R_2parameters}
   R_c(2)=-\frac{1}{4 N^2+1},
\end{equation}
which matches Eq.~(8.3a) of \cite{plunk2019direct} for the case $N = 1$.  The coefficient $Z_s(2)$ is free to be adjusted to satisfy other desired requirements. Note that additional Fourier modes may be retained to define a near-axis QI optimisation space, as done by Jorge et al (2022).

\subsection{Controlling Torsion}

\begin{figure}
\centering
  \includegraphics[width=0.49\textwidth]{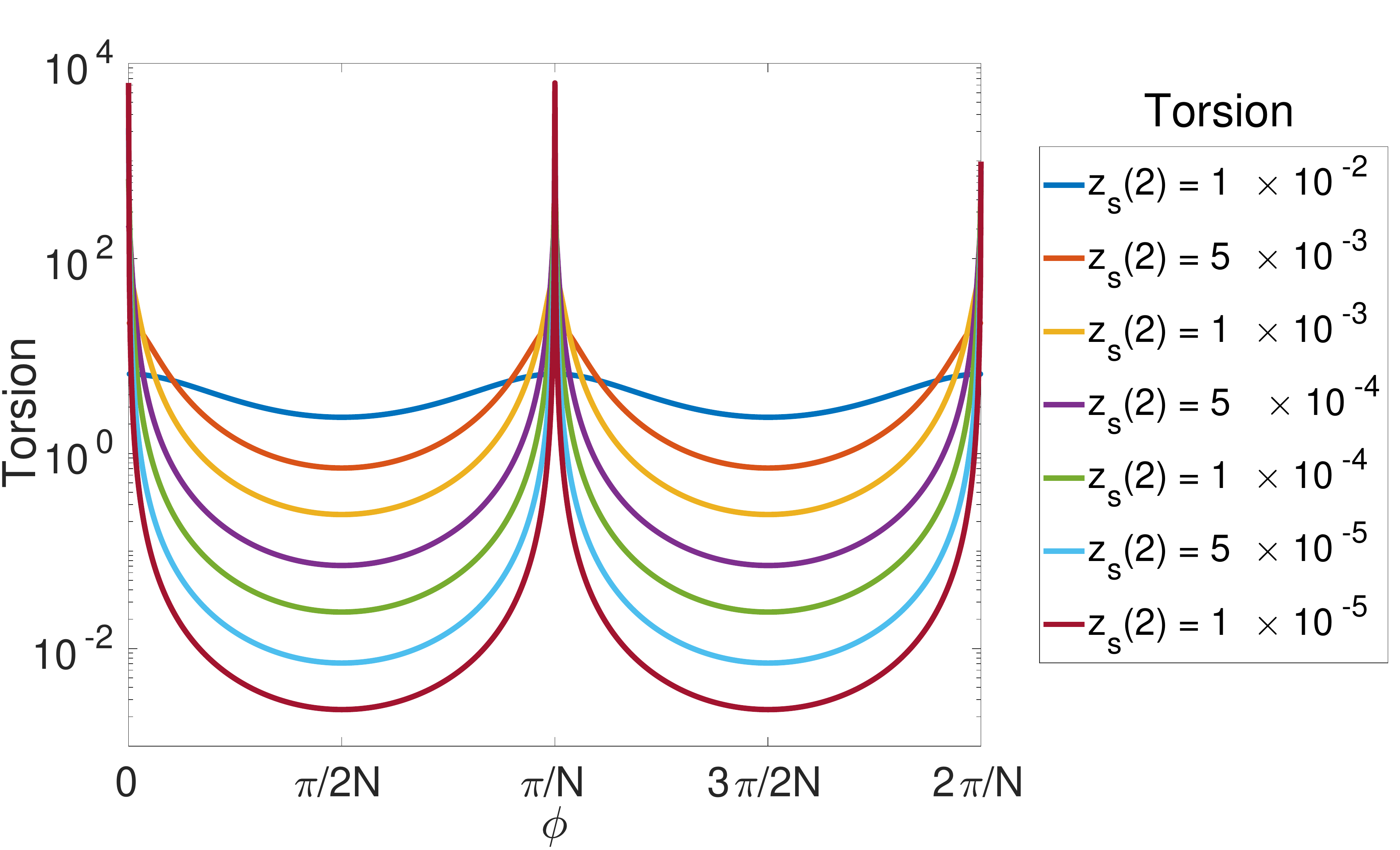}
   \includegraphics[width=0.49\textwidth]{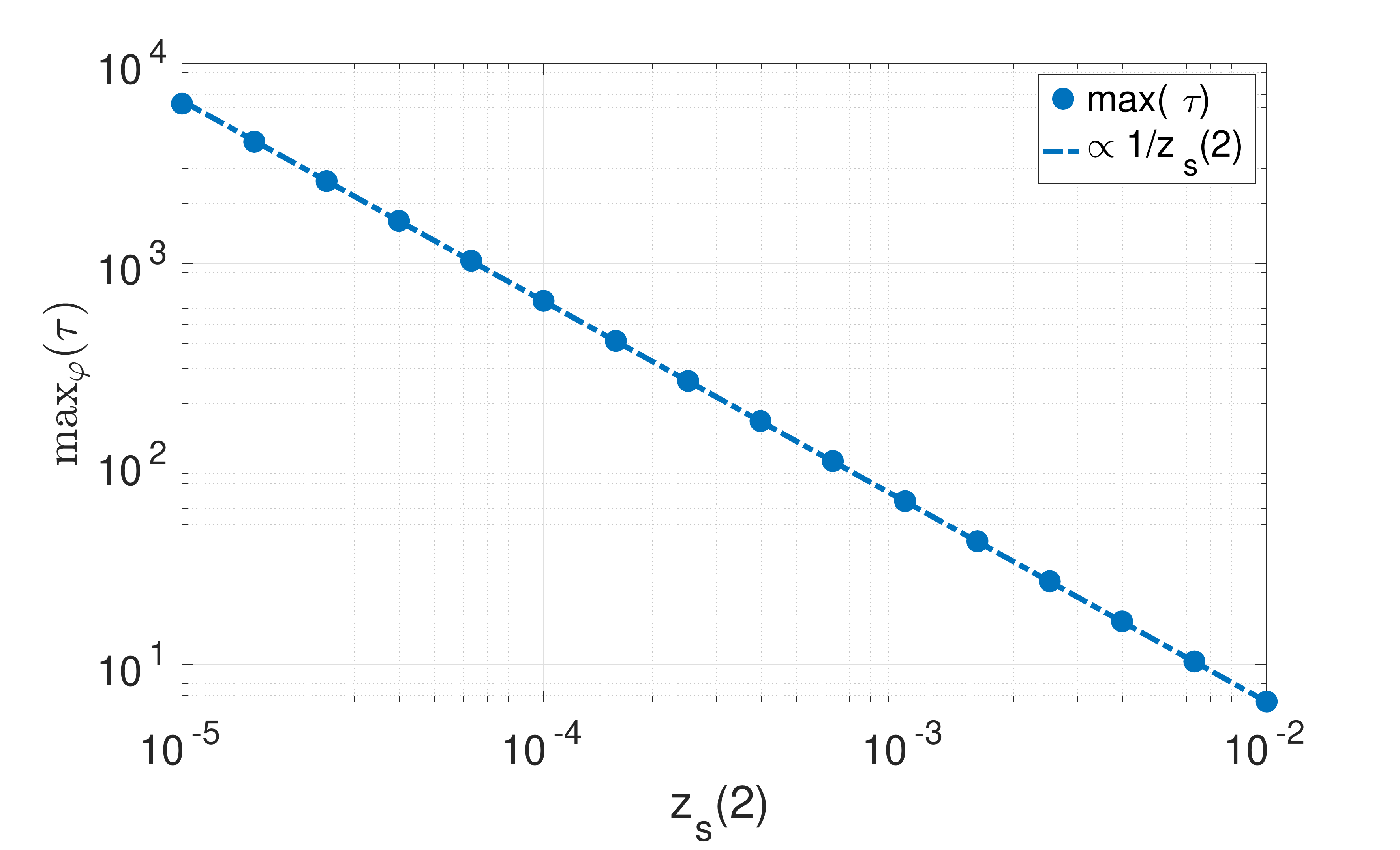} 
\caption{Torsion behaviour with decreasing value of $z_{s}(2)$. (left) Singularity in torsion at points of stellarator symmetry as $z_{s}(2)$ approaches zero. (right) Maximum torsion at different values of $z_{s}(2)$.   }
\label{fig:torsion-scan}
\end{figure}

Noting that a curve of zero torsion lies within a plane, it would seem straightforward to realize a stellarator-symmetric axis shape of low torsion by simply reducing the magnitude of its $z$ component, thereby constraining the curve to lie close to the $x$-$y$ plane.  We can do this with the single parameter curve defined by Eqs.~(\ref{eq:R_2param})-(\ref{eq:z_2param}), by letting the parameter $z_s(2)$ tend to zero.  Unfortunately this limit is not well-behaved, as shown in Figure \ref{fig:torsion-scan}.  Although torsion goes to zero almost everywhere, it tends to infinity in the neighbourhood of the zeros of curvature. 

\begin{figure}
\centering
  \includegraphics[width=0.49\textwidth]{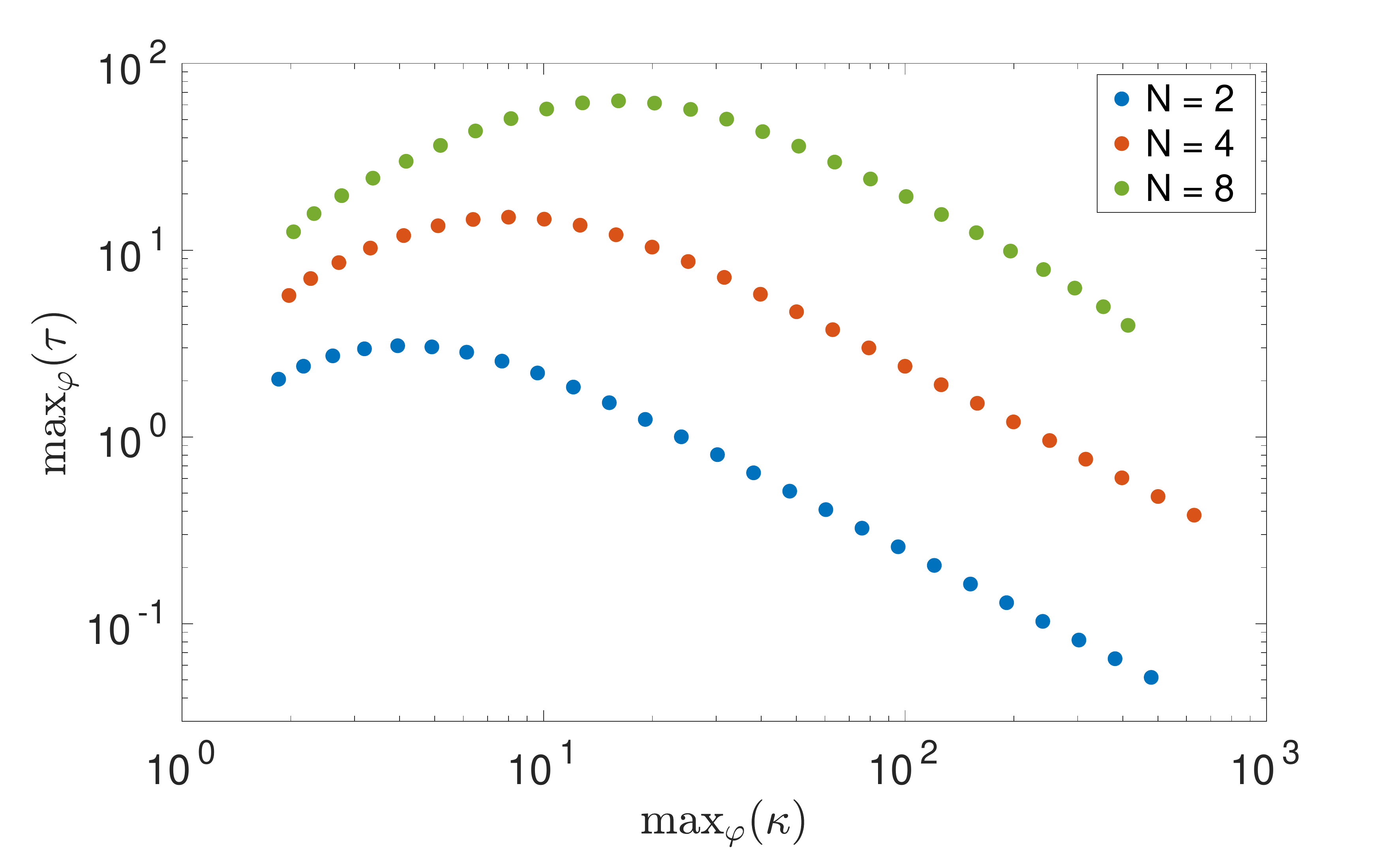}
   \includegraphics[width=0.49\textwidth]{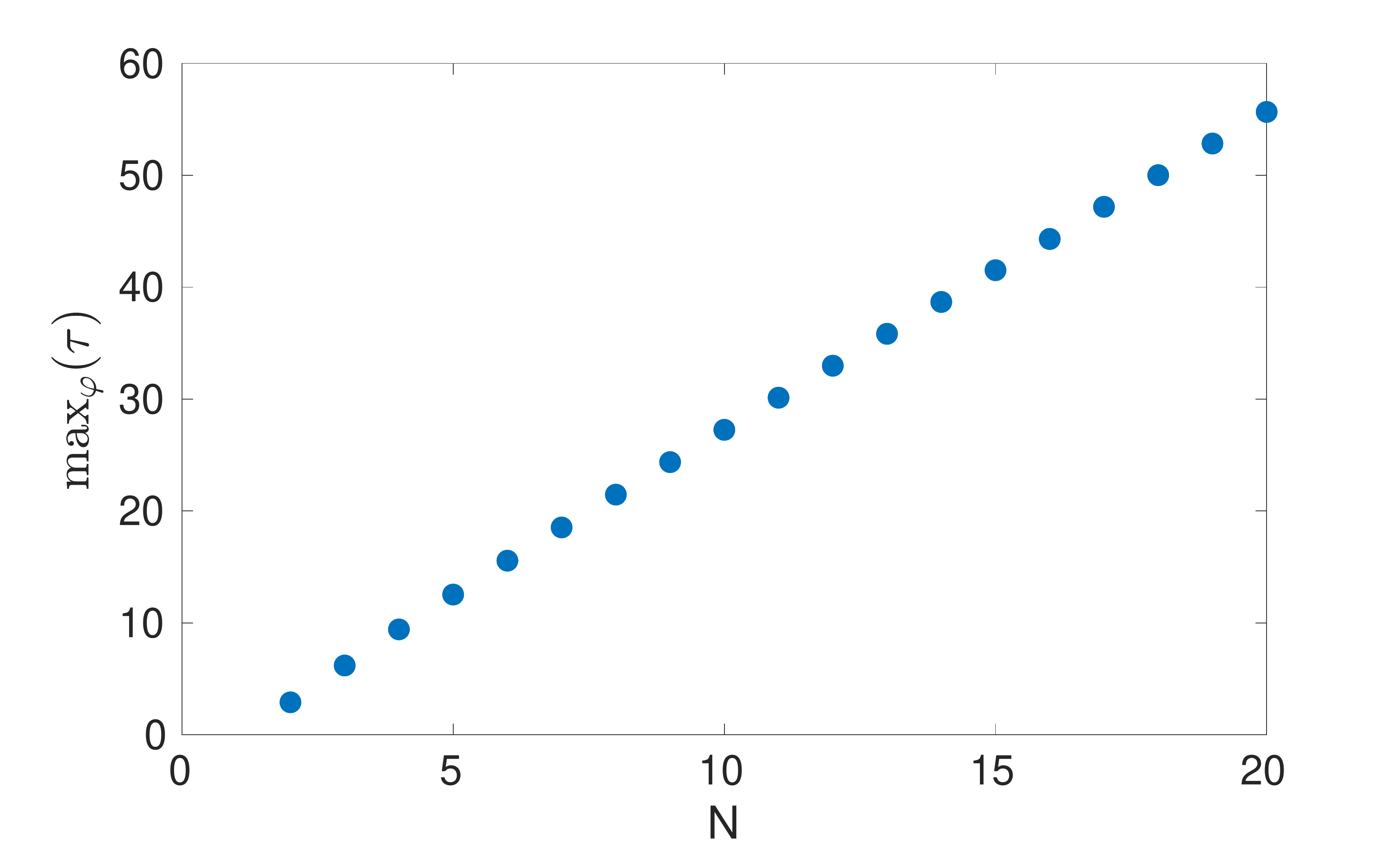} 
\caption{(left) Maximum torsion vs maximum curvature for $N=2,4,8$ (blue, orange, green). Curvature and torsion are individually maximized over $\phi$ for fixed values of $z_{s}(2)$. Values used correspond to large-$z_{s}(2)$ regime, $z_{s}(2)\gtrsim 0.4/N^2$. (right) Maximum torsion versus N for $z_{s}(2)$ chosen such that $\mathrm{max}_{\varphi}(\kappa) = 3$.}
\label{fig:tauVskappa}
\end{figure}

\begin{figure}
\centering
  \includegraphics[width=0.49\textwidth]{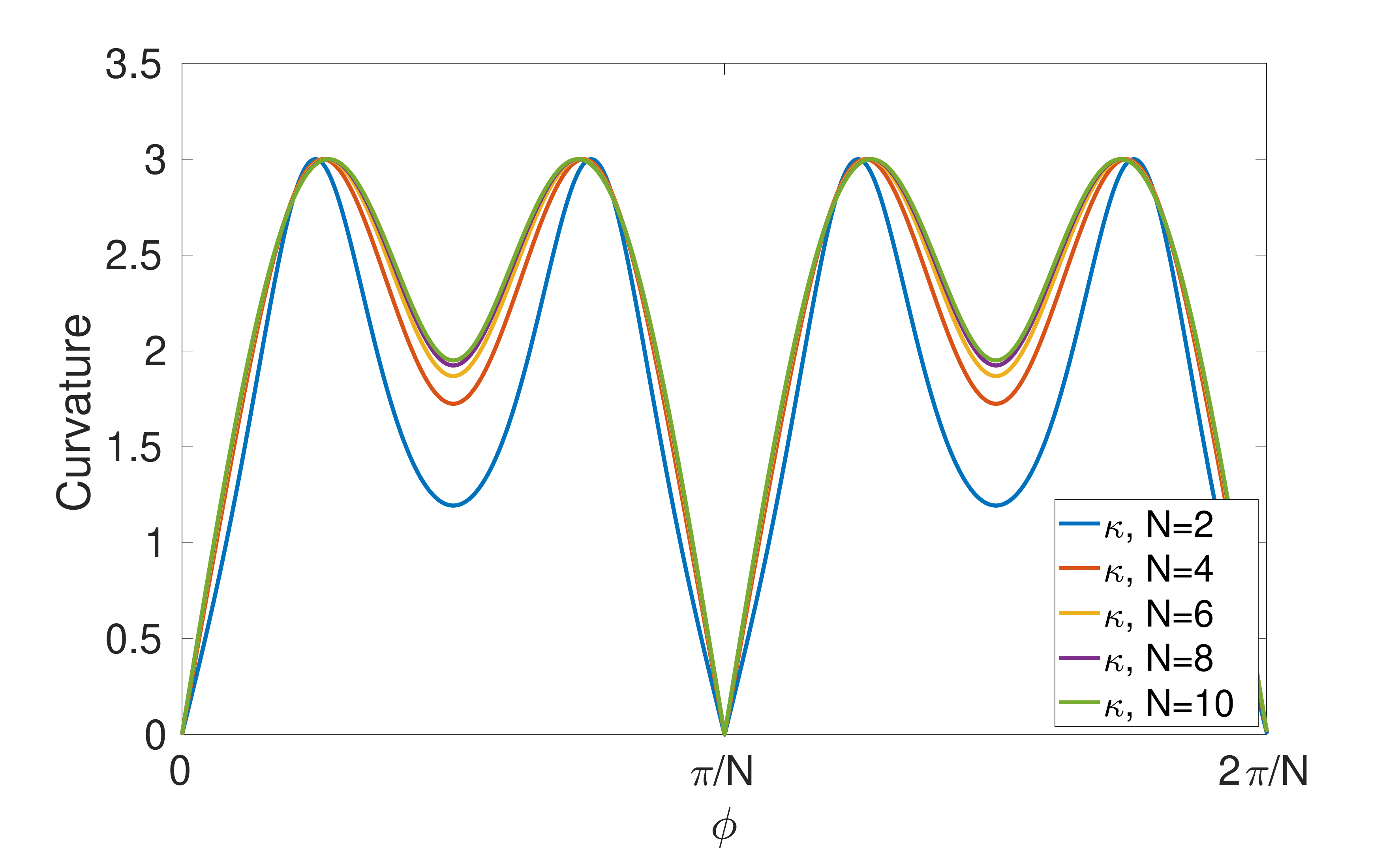}
   \includegraphics[width=0.49\textwidth]{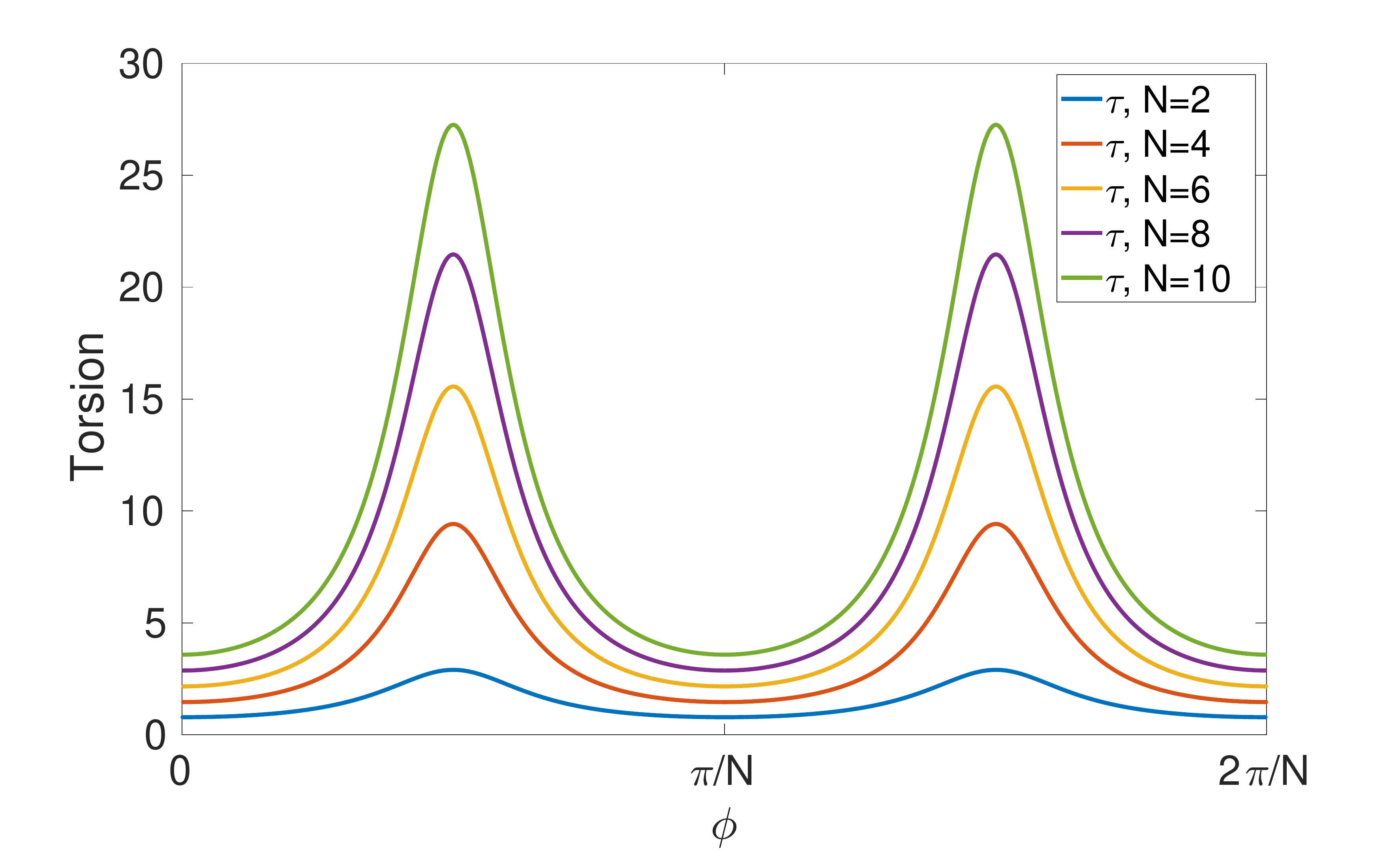} 
\caption{Curvature (left) and torsion (right) versus $\phi$ for different field period numbers, $N=2,4,6,8,10$. $z_{s}(2)$ is chosen such that $\mathrm{max}_{\varphi}(\kappa) = 3$.}
\label{fig:kappa_tau_N}
\end{figure}

Another approach to minimizing torsion is, somewhat paradoxically, to take the limit of large $z_s(2)$.  Such elongated axis shapes are nearly planar around the points of zero curvature, and have their torsion peaked midway between these points.  The torsion is, however, small in magnitude due to the large values of curvature around such points.  As figure \ref{fig:tauVskappa} confirms, the lowering of torsion by this method is accomplished only at the cost of raising the maximum value of curvature.  Furthermore, the maximum value of torsion obtained at a fixed value of maximum curvature ($\mathrm{max}(\kappa) = 3$) grows linearly with field period number, as shown in the second panel of figure \ref{fig:tauVskappa}. Requiring the maximum value of curvature to remain bounded  when increasing the number of field periods results in the maximum of torsion increasing, as shown in figure \ref{fig:kappa_tau_N}. Finding closed curves with torsion and curvature remaining under a certain value gets more difficult when increasing the number of field periods. This might be a reason why finding good solutions with $N>1$ is challenging for the optimisation procedure described in \cite{Jorge2022}.

\section{QI construction, two-field-period example}\label{example_2FP}

\begin{figure}
    \centering
    \includegraphics[width=0.6\textwidth]{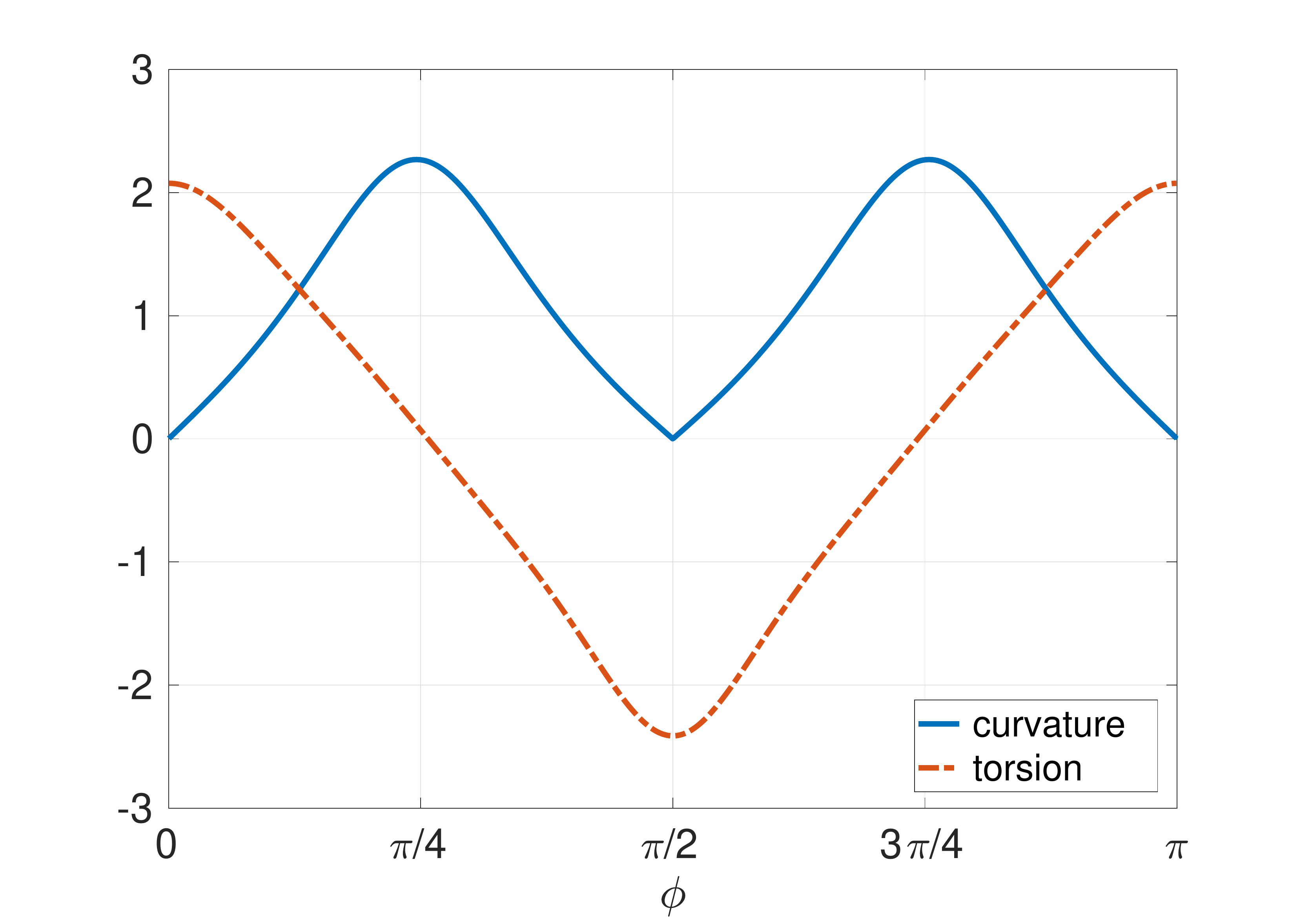}
    \caption{Curvature (unsigned) and torsion profiles for one field period of the axis described by (\ref{eq:R0_2FP}) and (\ref{eq:Z0_2FP}). The  curvature has zeros at points of extrema of the on-axis magnetic field $B_{0}$ as required by the theory.}
    \label{fig:kappa_tau_2FP}
\end{figure}

The construction of a two-field-period configuration will now be described. The axis shape was chosen using equation (\ref{eq:R_2parameters}) for the case $N=2$, yielding
\begin{equation}\label{eq:R0_2FP}
    R = 1 - \frac{1}{17} \cos{(4\phi)},
\end{equation}
\begin{equation}\label{eq:Z0_2FP}
    z =  0.3921 \sin{(2\phi)} + 4.90\times10^{-3}\sin{(4\phi)}.
\end{equation}
The $z$-coefficients were chosen to limit the curvature and torsion to tolerable levels. The normal vector $\mathbf{n}^{s}$ does not complete any full rotation around the axis, hence $m=0$. This curve contains points of zero curvature as shown in figure \ref{fig:kappa_tau_2FP}. The location of these points coincide with the extrema of the on-axis magnetic field strength, which is chosen as 
\begin{equation*}
    B_{0} = 1 + 0.15\cos{(2\varphi)},
\end{equation*}

In general, the values of the toroidal coordinates $\varphi$, the Boozer angle, and $\phi$, the cylindrical angle, are not the same. However, thanks to stellarator symmetry, they coincide at the extrema points of $B_0(\varphi)$, which thus also correspond to the zeros of the axis curvature.

The choice of $d(\varphi)$ has an important effect on the elongation of the plasma boundary as can be seen from Eq.~(\ref{eq:X0_Y0}). We observed that keeping it proportional to $\kappa^{s}(\varphi)$ helped reducing the elongation to manageable levels. For this example it was chosen as $d(\varphi) = 1.03~ \kappa^{s} $. The parameter $k$ entering Eq.~(\ref{eq:alpha_final}) controls the deviation from omnigenity and was set to $k=2$. 

To find numerical solutions, we first find the signed Frenet-Serret frame quantities of the axis; $\kappa^s$, $\tau$, and $(\mathbf{t,n^{s},b^{s}})$. Then, the relation $\varphi(\phi)$ as well as $G_{0}$, need to be found along the axis. This is done by iteratively solving equations (8.1) of \cite{plunk2019direct}
\begin{align}
\frac{d\varphi}{d\phi}= \frac{B_0}{|G_0|}\left|\frac{d \vect{r}}{d \phi} \right|,
\hspace{0.3in}
|G_0| = \frac{1}{2\pi N}\int_0^{2\pi}d\phi \;B_0(\varphi(\phi))\left|\frac{d \vect{r}}{d \phi} \right|.
\end{align}
We then proceed to solve equation (\ref{eq:sigma}), self-consistently with $\iotaslash$, for one field period, i.e. in the region $\varphi \in  [0,2\pi/N]$.  

\begin{figure}
    \centering
    \includegraphics[width=0.95\textwidth]{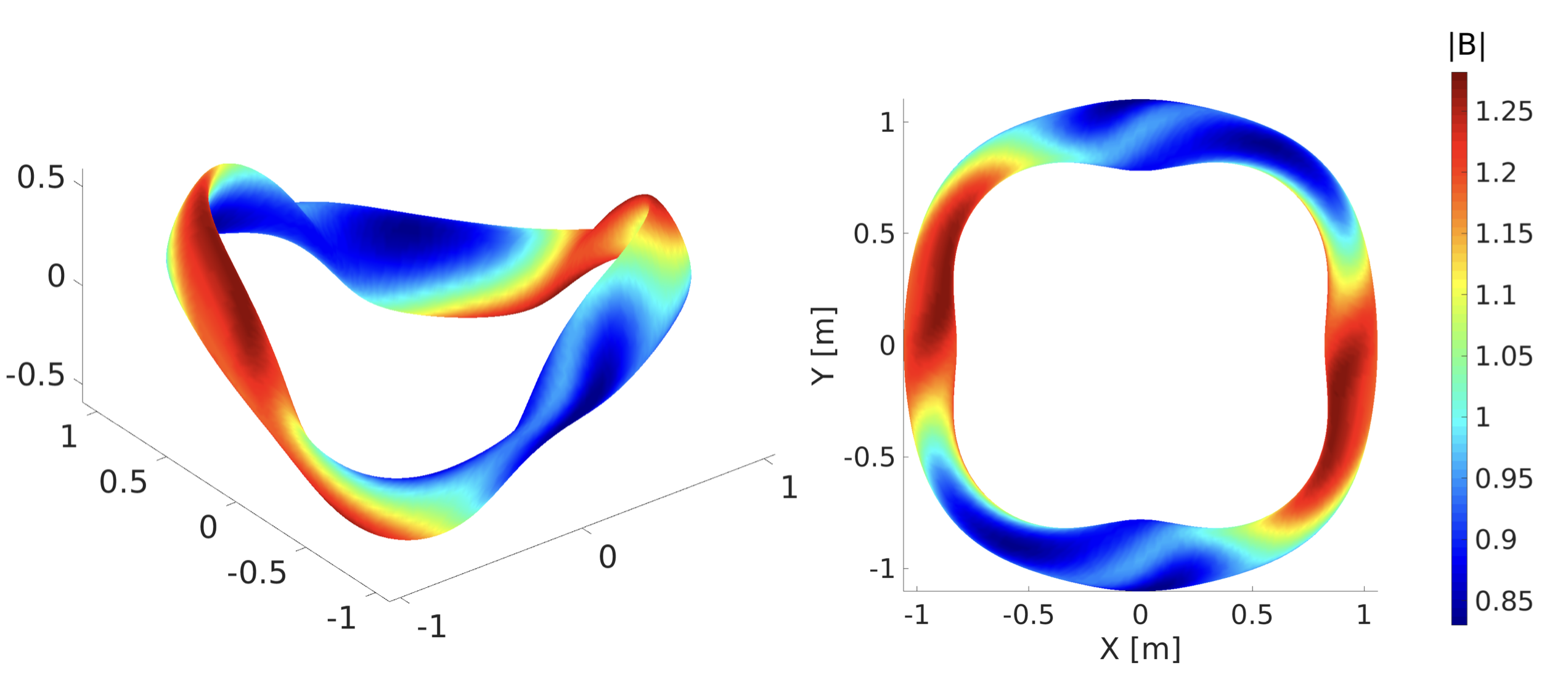}
    \caption{Intensity of the magnetic field on the plasma boundary for a 2 field period, A=10 configuration. Side view (left) and top view (right) are shown.}
    \label{fig:Boundary_2FP}
\end{figure}

\begin{figure}
\centering
  \includegraphics[width=0.49\textwidth]{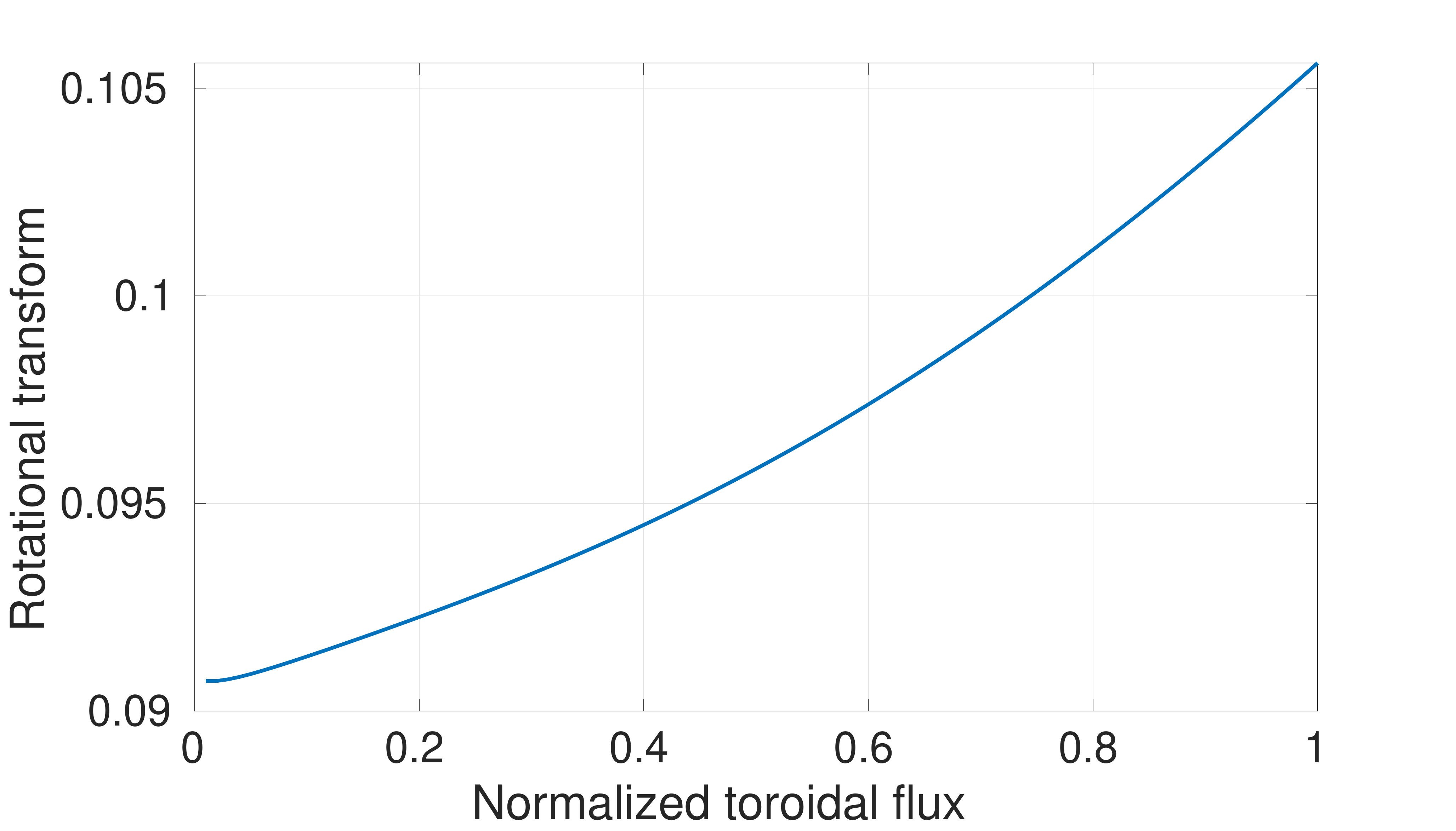}
  \includegraphics[width=0.49\textwidth]{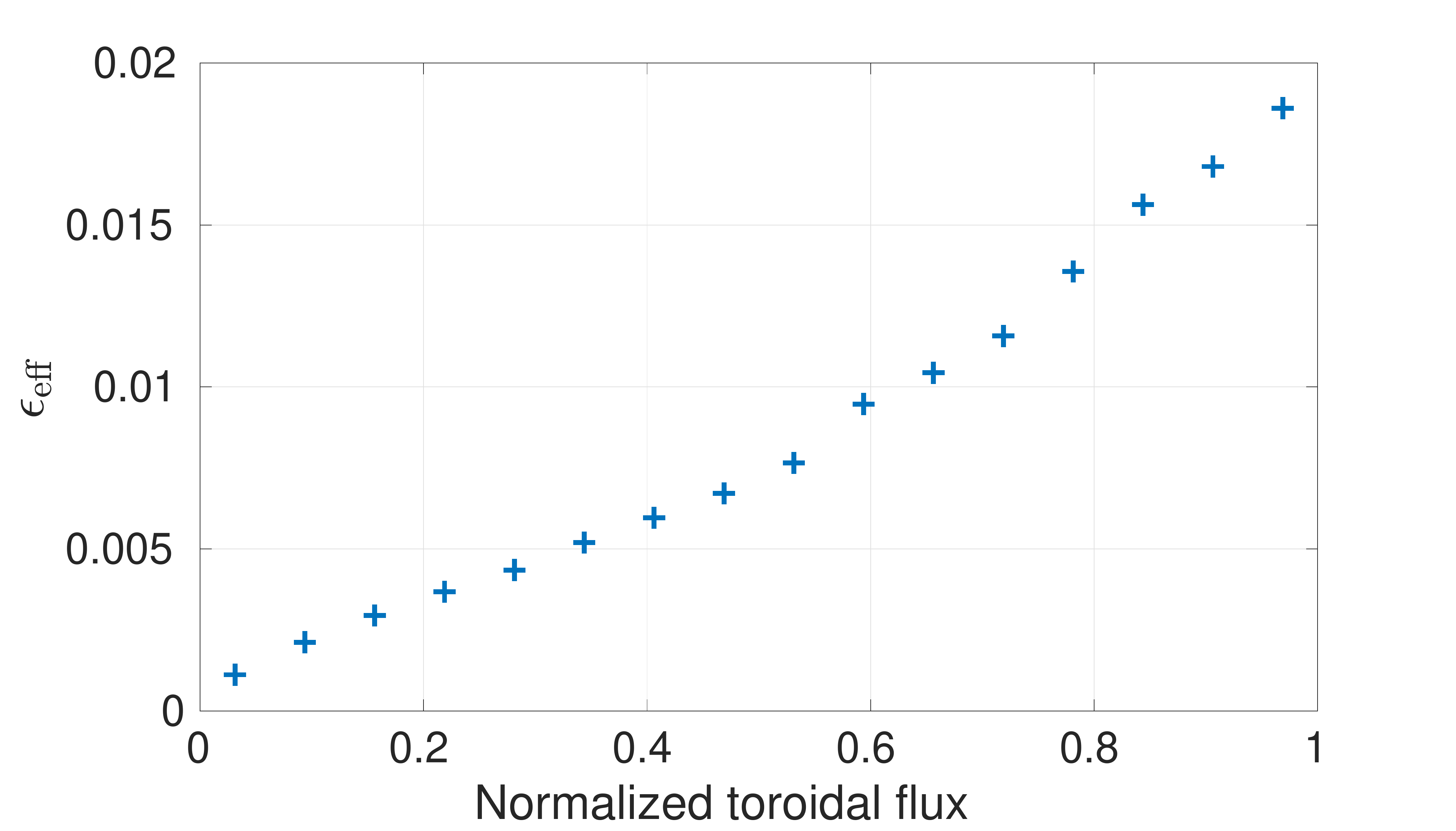}
\caption{ (left) Rotational transform, $\iotaslash$, and  (right) effective ripple, $\epsilon_{\mathrm{eff}}$} for an $N=2$, $A=10$ configuration.
\label{fig:iota_2FP_eps_eff}
\end{figure}

A boundary is constructed with aspect ratio $A=10$, where $A$ can be expressed in terms of the distance from the axis as 
\begin{equation} A=\sqrt{\frac{\bar{B_{0}}}{2}}\frac{R_c(0)}{\epsilon}, 
\end{equation}
where $\bar{B_0}$ is the average value of $B_{0}$. This boundary is  then used to find a fixed-boundary magnetic equilibrium with the code VMEC. The strength of the magnetic field on the boundary is shown in figure \ref{fig:Boundary_2FP}. The rotational transform profile obtained with VMEC is shown in figure \ref{fig:iota_2FP_eps_eff} and coincides with the value calculated numerically from Eq.~(\ref{eq:sigma}) $\iotaslash = 0.107$. The maximum elongation of the flux surface cross-sections, as defined by Eq.~(\ref{eq:elongation}) is $e_{\mathrm{max}}=4.4$.

The effective ripple, $\epsilon_{\mathrm{eff}}$, is a  simple and convenient parameter that characterizes low-collisionality neoclassical transport of electrons \citep{Beidler-2011}. We calculate it using the procedure described by \cite{drevlak2003effective} in 16 radial points, and find an $\epsilon_{\mathrm{eff}}$ below 1\% up to mid-radius and lower than 2\% everywhere in the plasma volume, see Figure \ref{fig:iota_2FP_eps_eff}. 

\begin{figure}
\centering
  \includegraphics[width=0.49\textwidth]{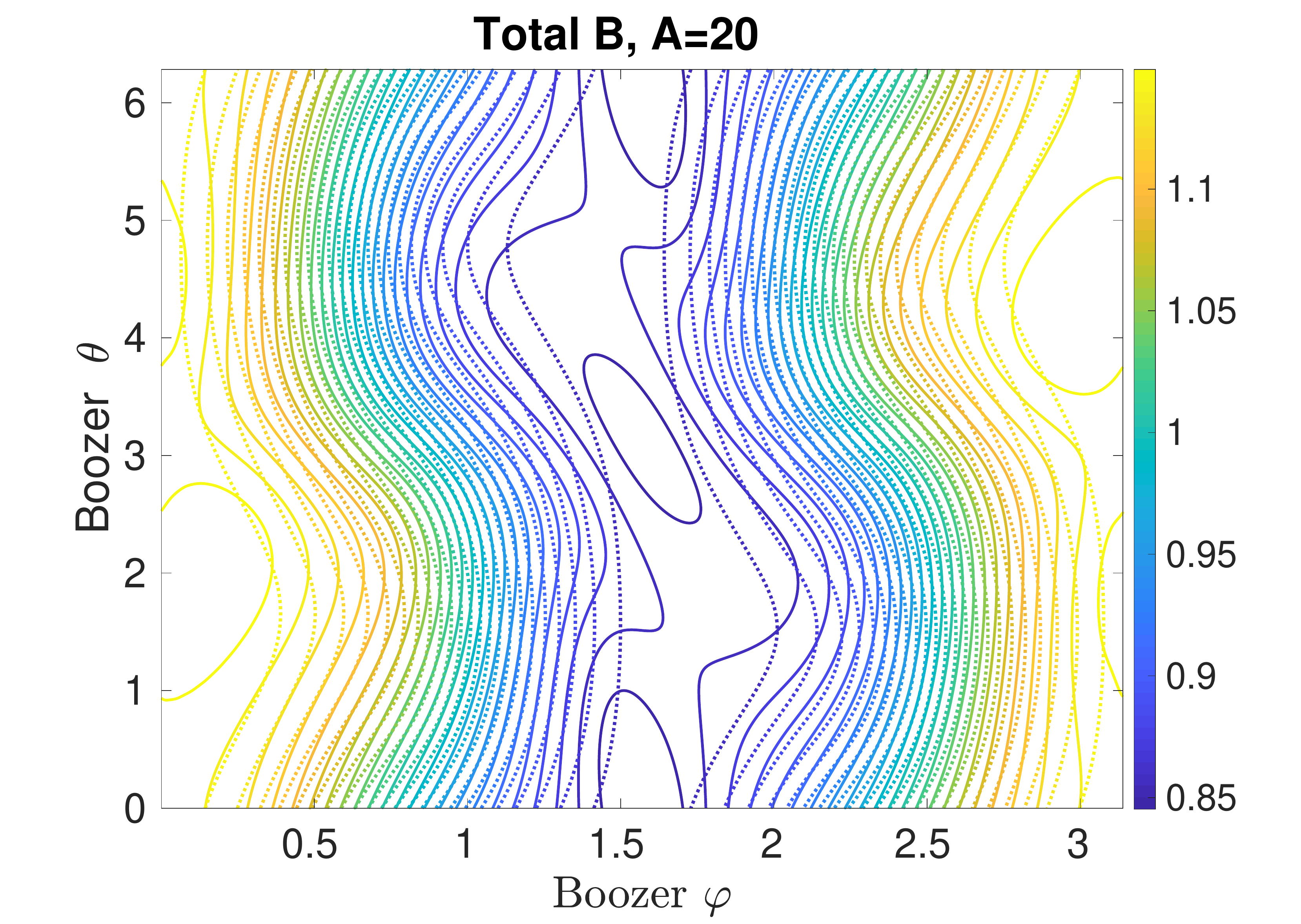}
  \includegraphics[width=0.49\textwidth]{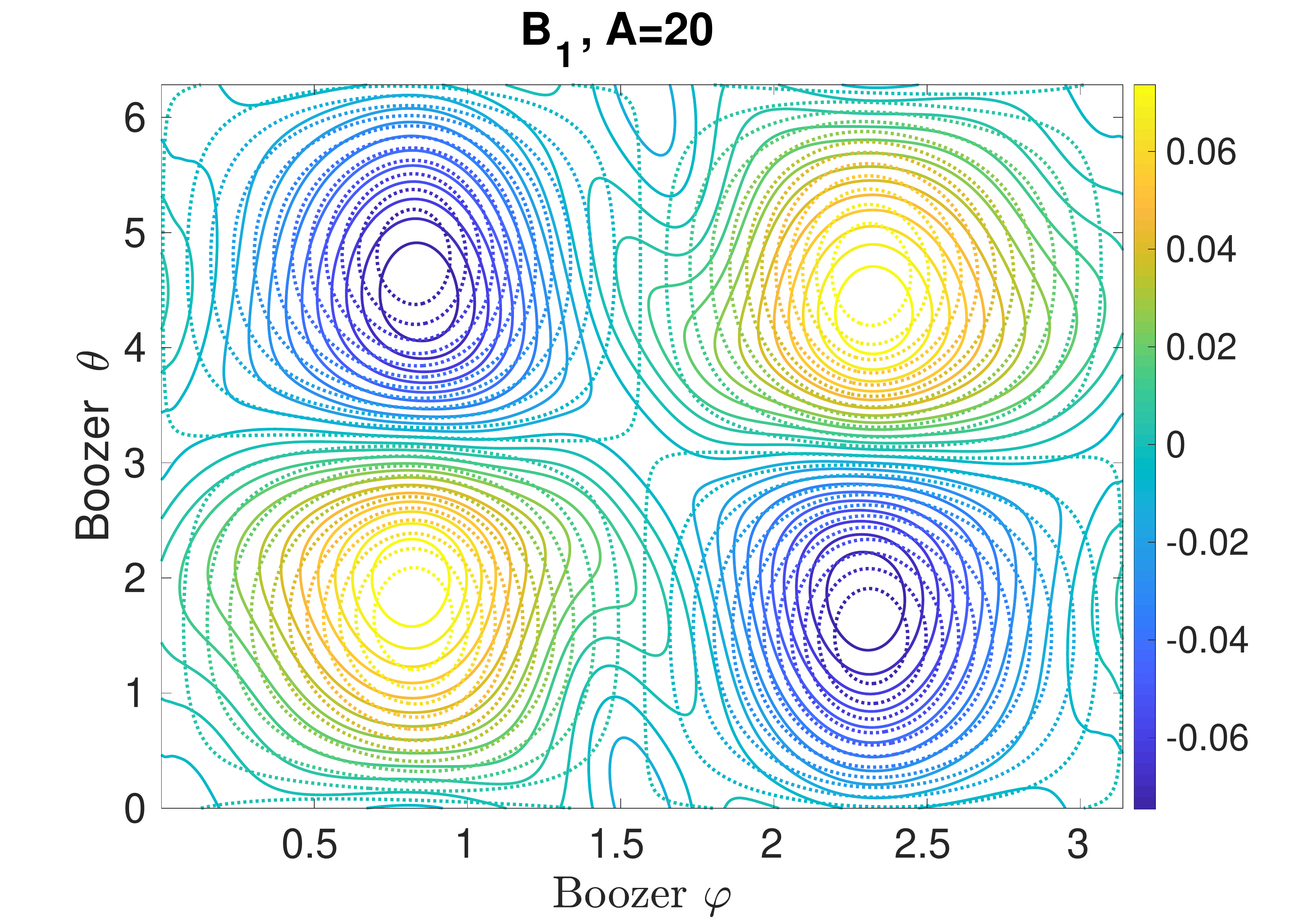}
  \includegraphics[width=0.49\textwidth]{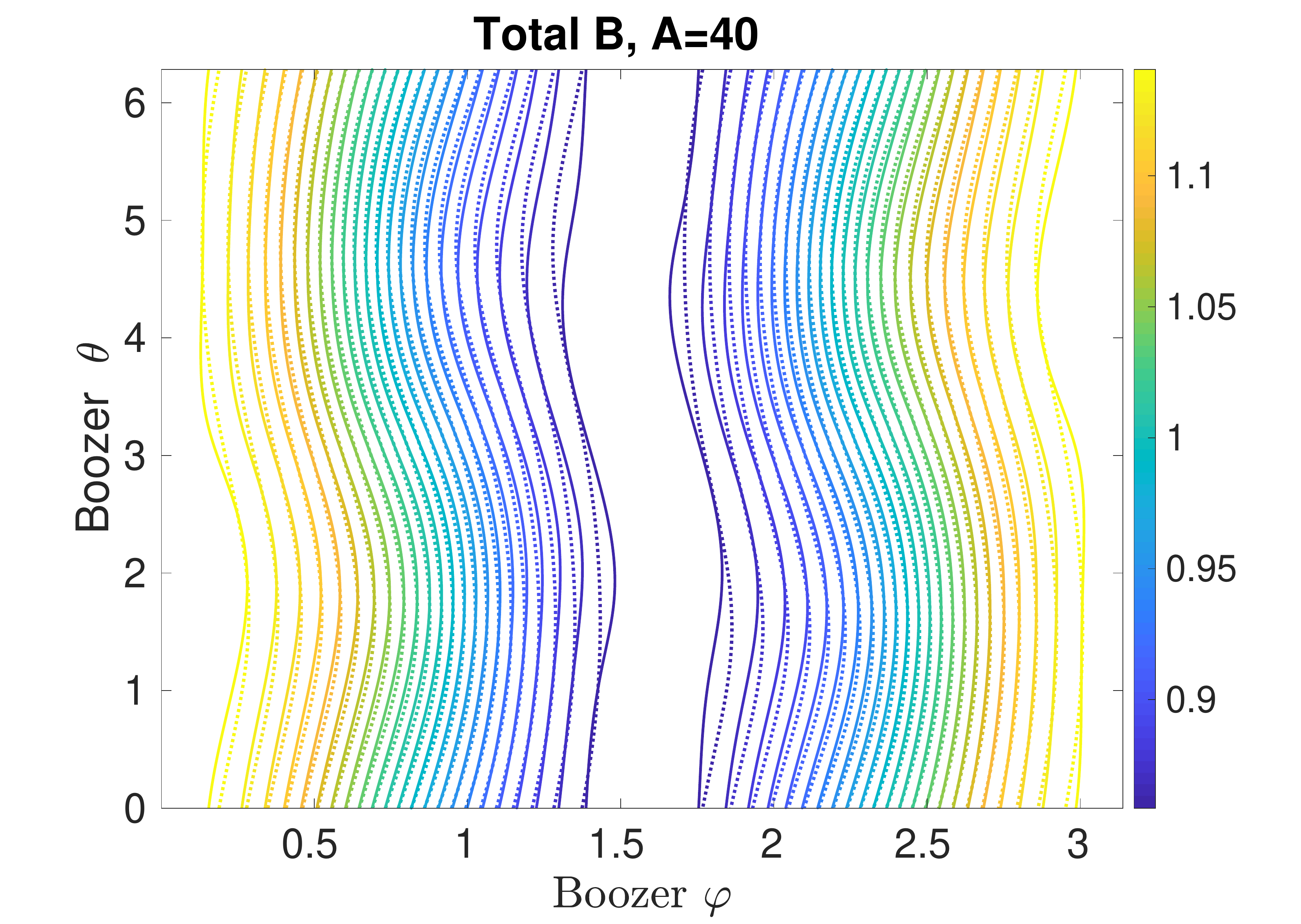}
  \includegraphics[width=0.49\textwidth]{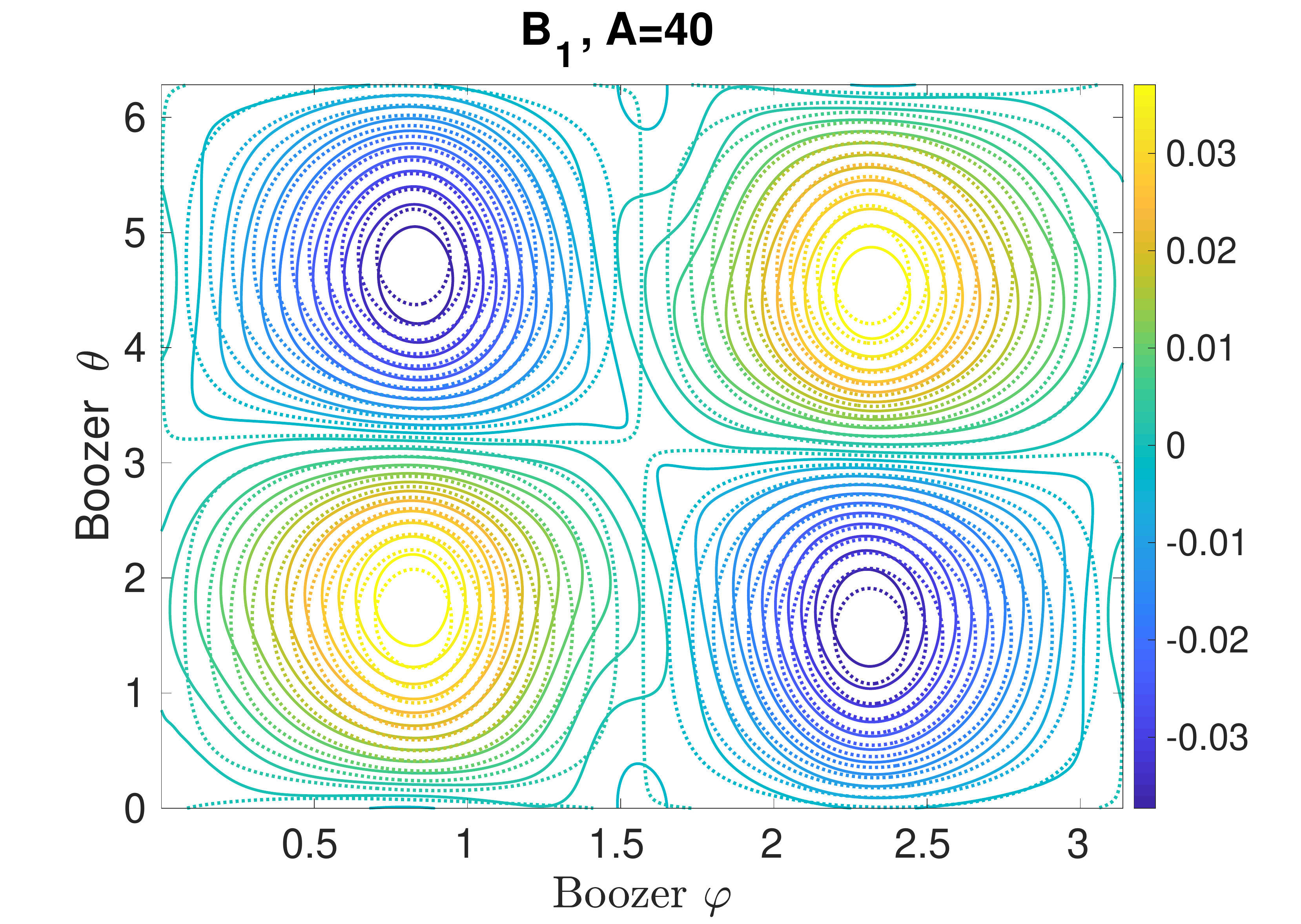}
  \includegraphics[width=0.49\textwidth]{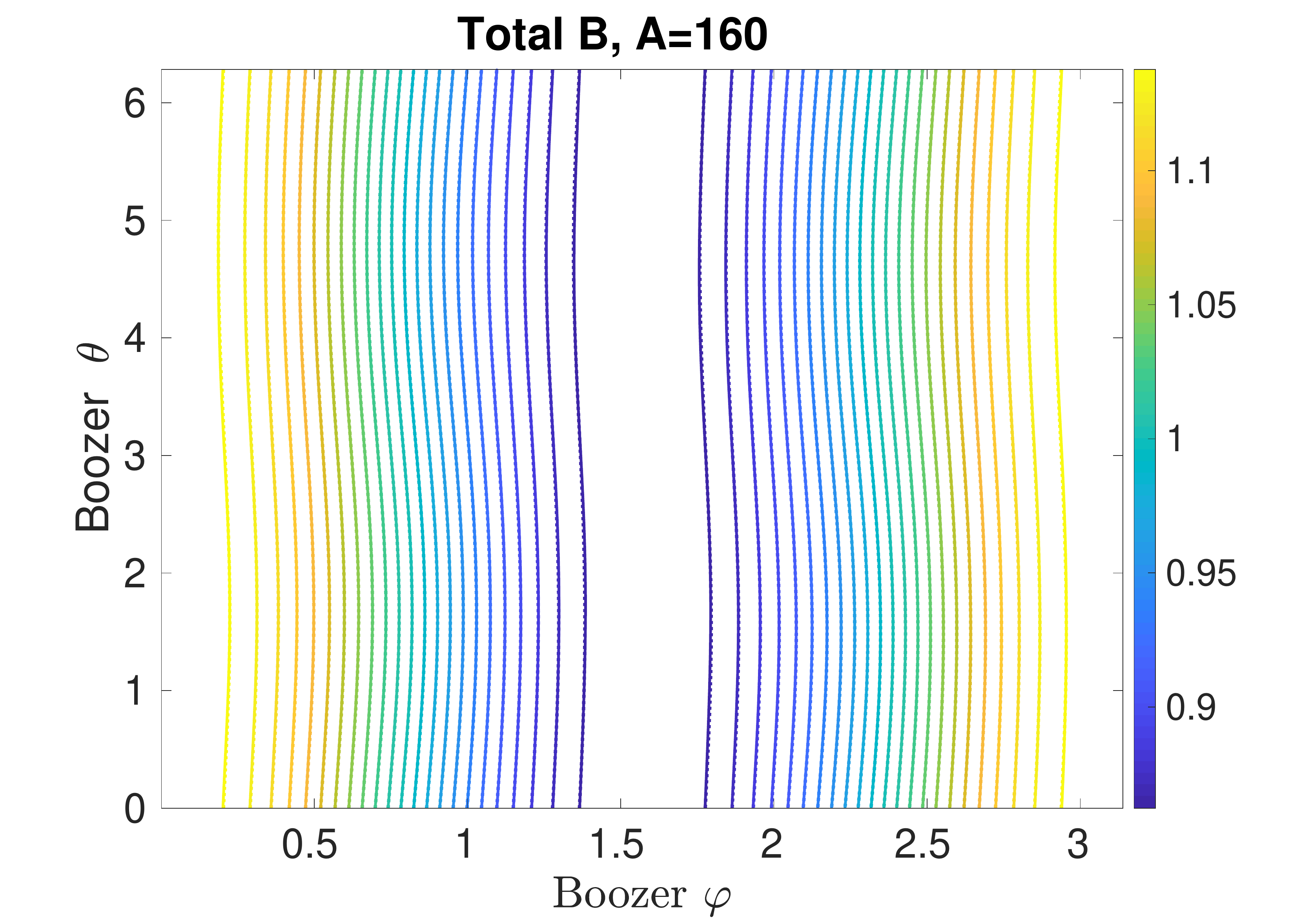}
  \includegraphics[width=0.49\textwidth]{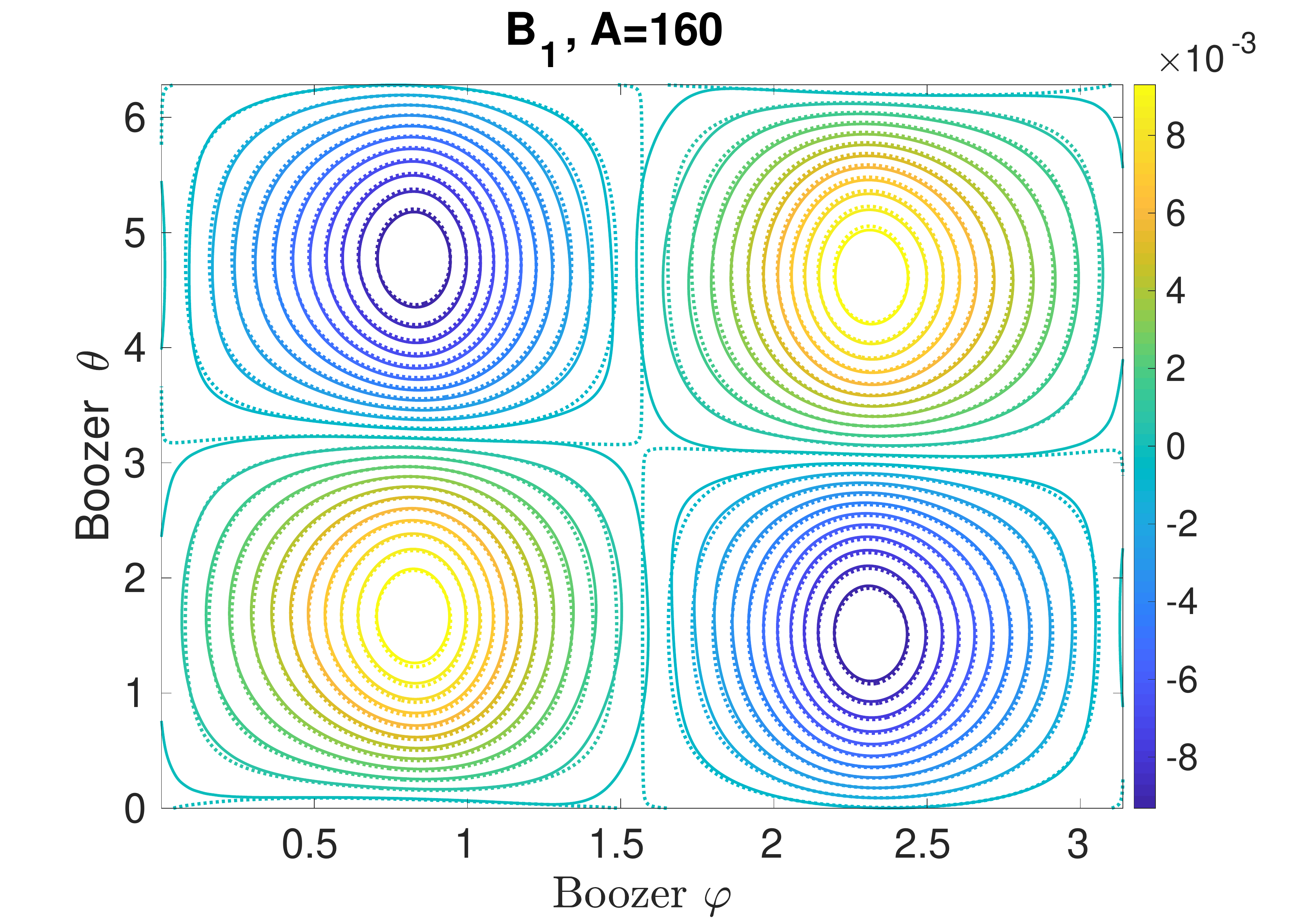}
\caption{Contours of the magnetic field intensity on the plasma boundary, $B$, are shown for increasing values of aspect ratio, starting at $A=20$ on the top. Dotted lines are the contours obtained from the construction and solid lines from VMEC and BOOZ\_XFORM. Contours on the right column correspond to the first order correction to $B$ and left column the total magnetic field.}
\label{fig:BContours_Comp_2FP}
\end{figure}

\begin{figure}
    \centering
    \includegraphics[width=0.6\textwidth]{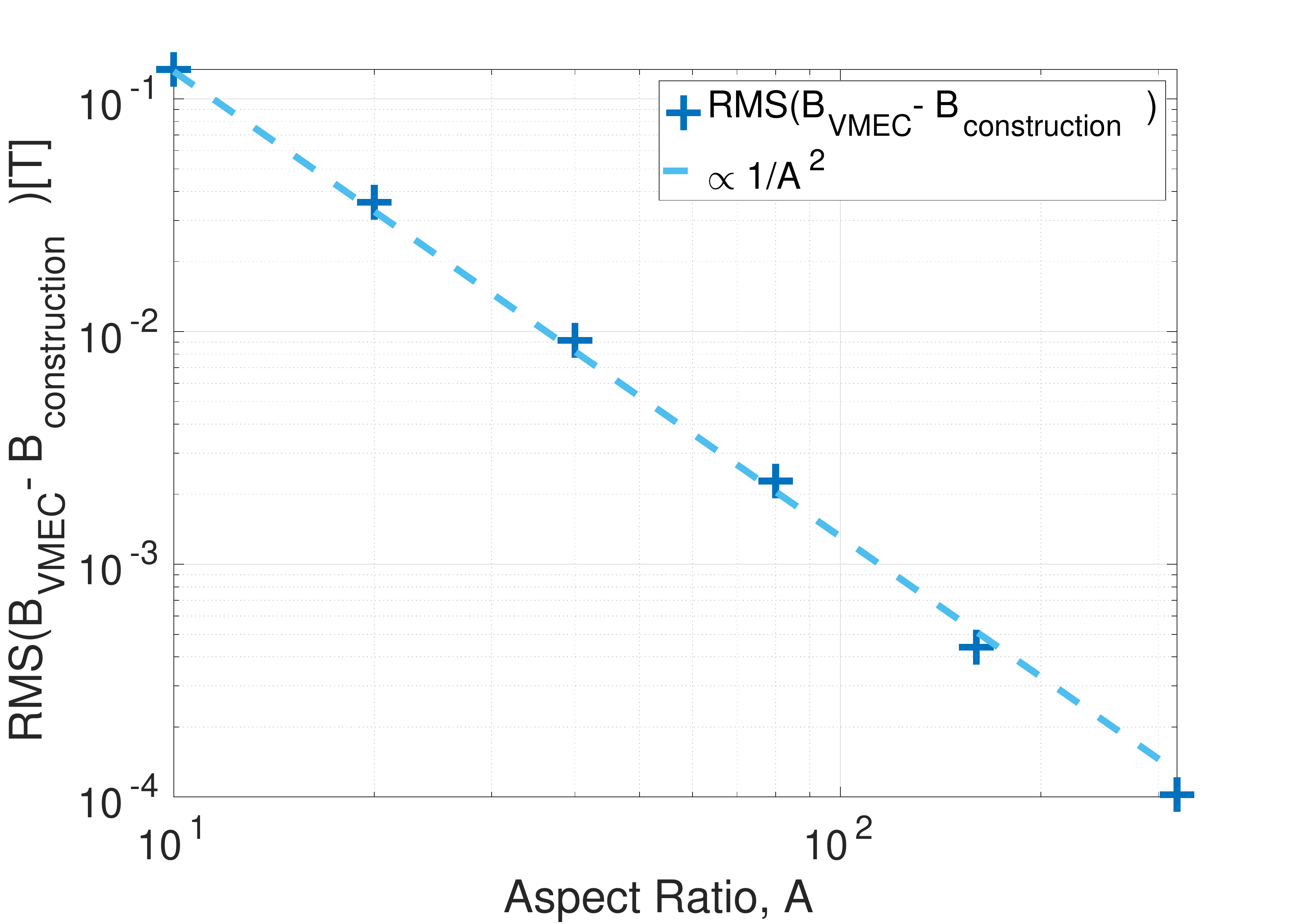}
    \caption{Root-mean-squared difference between the magnetic field from the construction and that calculated by VMEC, for different aspect ratios. The dashed line shows the expected scaling $\propto 1/A^{2}$. }
    \label{fig:B_rms_2FP}
\end{figure}

The boundary shape was generated for different aspect ratios, i.e. different distances from the axis. A comparison between the contours of constant magnetic field strength obtained from the near-axis expansion, Eq.~(\ref{eq:magneticField}), and those obtained from the VMEC calculation and transformed to boozer coordinates using BOOZ\_XFORM (\cite{sanchez2000ballooning}) is shown in Fig.~\ref{fig:BContours_Comp_2FP}. The difference between the results decrease with increasing the aspect ratio, as expected since the expansion is performed in the distance from the axis. For the largest aspect ratio, 160, the difference is almost imperceptible. The root-mean-squared difference between both magnetic fields is calculated for each aspect ratio and the results are shown in figure~\ref{fig:B_rms_2FP}. The scaling with the aspect ratio is as expected from a first order expansion, proportional to $1/A^2$.  

\section{A family of constant torsion QI solutions}\label{Constant_Torsion_QI}

As illustrated by the previous example, we are capable of directly constructing QI equilibria with low electron neoclassical transport, as measured by $\epsilon_{\mathrm{eff}}$, at aspect ratios comparable to existing devices. Performing an optimisation procedure in the space of QI solutions described by the near-axis expansion has led to the discovery of $N{=}1$ configurations with excellent confinement properties as shown in \cite{Jorge2022}. But finding configurations with more than one field period and good confinement has proved challenging, even when using optimisation procedures. In order to explore the role of the axis shape has in causing this behaviour, we choose a family of axis shapes with $N{=}2,3,4$, constant torsion, and equal per-field-period maximum curvature. Constant torsion was chosen for simplicity in order to obtain smooth solutions for $\sigma$ from equation (\ref{eq:sigma}). 

The axis shapes chosen are closed curves with minimal bending energy and constant torsion as described in \cite{pfefferle2018non}. Their curvature and torsion are
\begin{equation}\label{eq:tau}
   \tau = 4 X(p)K(p) N/L,
\end{equation}
\begin{equation}\label{eq:kappa}
 \kappa^{2}(s)=\kappa_{0}^{2}\left[ 1 - \mathrm{sn}^{2}\left(\frac{\kappa_{0}}{2p}(s-0.5L),p\right)\right],
\end{equation}
where $\mathrm{sn}(a,b)$ is the Jacobi elliptic sine function, $L$ the curve length, $N$ the number of field periods, $p$ a parameter yet to be chosen, and $\kappa_{0}$ the maximum curvature. $K(p)$ is the complete elliptical integral of first kind and $X^{2}(p)= 2E(p)/K(p) -1$, where $E(p) = E(\pi/2,p)$ is the incomplete elliptic integral of the second kind. 

\begin{figure}
    \centering
    \includegraphics[width=0.49\textwidth]{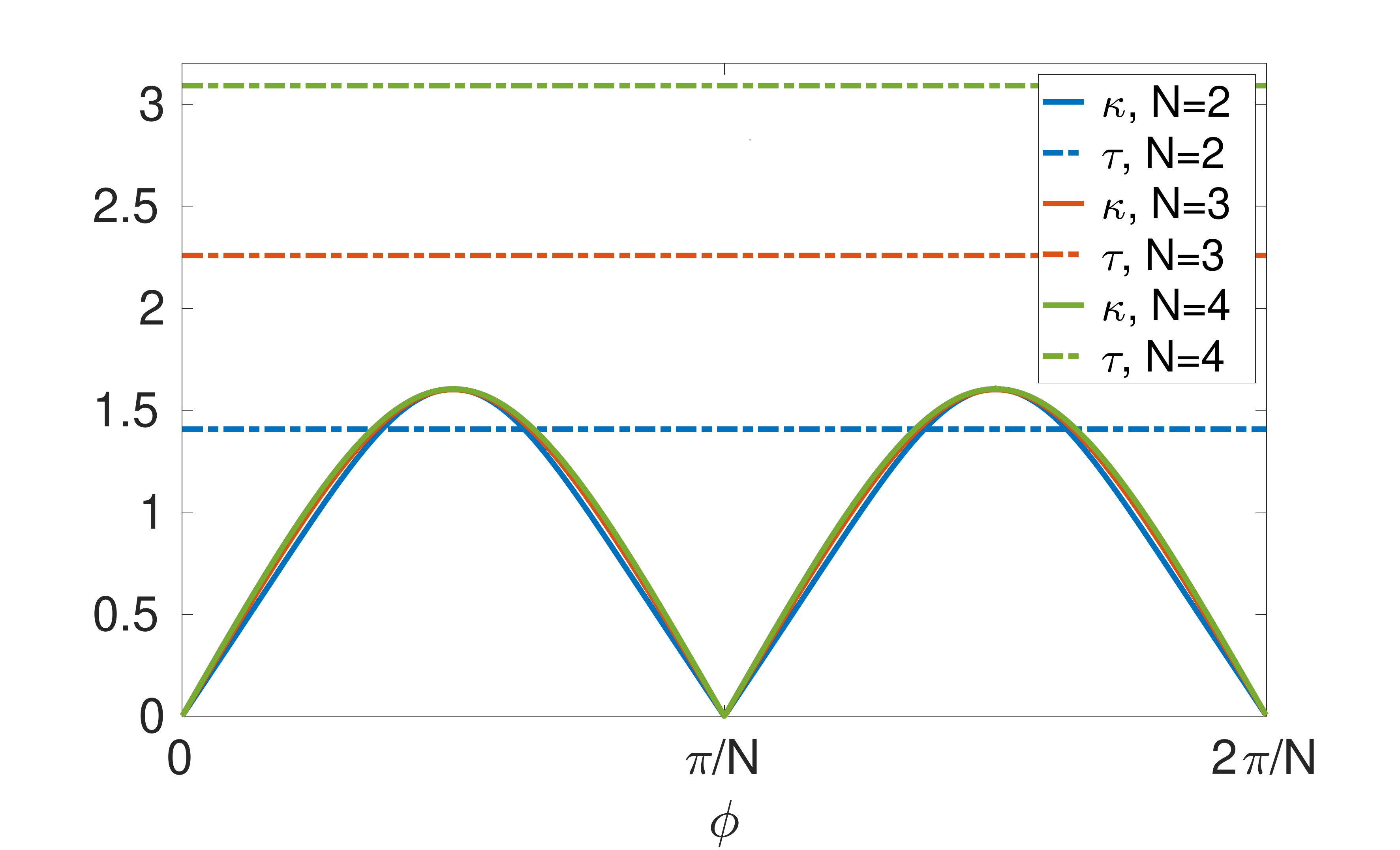}
    \includegraphics[width=0.49\textwidth]{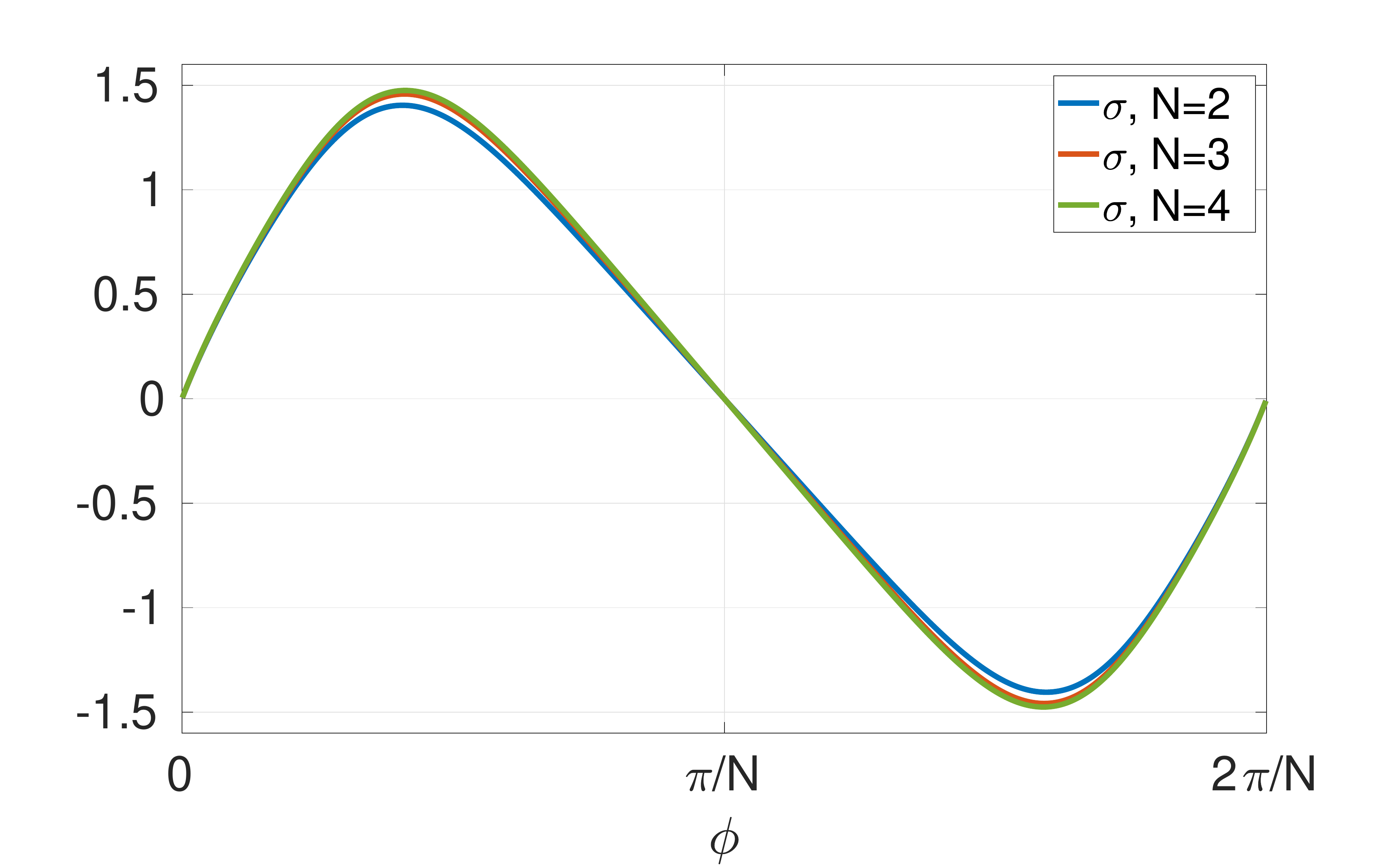}
    \caption{(left) Curvature, $\kappa$, and torsion, $\tau$, with respect to the toroidal angle $\phi$ of the magnetic axes used in the construction of the different N field-period solutions. $\kappa(\varphi)$ and $\tau(\varphi)$ are given by equations (\ref{eq:kappa}) and (\ref{eq:tau}). The maximum curvature, $\kappa_0 = 1.6$, is the same for the three cases. (right) $\sigma$ profiles obtained by solving equation \ref{eq:sigma} for each of the curves described by $\kappa$ and $\tau$ on the left. }
    \label{fig:kappa_tau_Pfefferle}
\end{figure}

The maximum curvature is chosen as $\kappa_0{=}1.6$ for all cases. Given a number of field periods $N$, the parameter $p$ is scanned until a closed curve is found. We find $p_{2}{=}0.4527$, $p_{3}{=}0.5927$ and $p_{4}{=}0.6580$, for two, three and four field periods, respectively.    
The curvature per field period has the same maximum and similar toroidal dependence for all three cases and has zeros at multiples of $\pi/N$, as required by the construction. The necessary torsion for a closed curve increases with $N$ as seen in figure (\ref{fig:kappa_tau_Pfefferle}). The Frenet-Serret formulas (Eqs. \ref{eq:Frenet-Serret formulas}) are then used to find the curve described by these values of $\kappa$ and $\tau$, as described in section \ref{Axis_Shape}. For the three curves $m=1$.

Using these curves as axis shapes, three configurations were constructed following the method described in section \ref{example_2FP}. All per-period parameters and functions used in the construction are kept the same as in section \ref{example_2FP}; $d(\varphi) = 1.03\kappa^{s} $, the parameter $k$ entering Eq.~(\ref{eq:alpha_final}) was set as $k=2$, and the magnetic field on axis as
\begin{equation*}
   B_{0} = 1 + 0.15\cos{(N\varphi)}.
\end{equation*}
The main differences between these configurations is the value of the torsion $\tau$ and the number of field periods. Since we are solving equation (\ref{eq:sigma}) for $\sigma$ in a single period, we can compare the per field-period solutions with respect to a scaled angular variable $\varphi/N$, as seen in figure \ref{fig:kappa_tau_Pfefferle}(right). The solution for $\sigma$ in one field period for each of these cases is very similar and only deviates in those regions where the curvature values differ, as expected. Consequently, the values found for the per-period rotational transform, $\iotaslash_{N}=\iotaslash/N$, are also very similar; $\iotaslash_{\mathrm{2}}= 0.4565$, $\iotaslash_{\mathrm{3}}= 0.4674$,  $\iotaslash_{\mathrm{4}}= 0.4707$.    

\begin{figure}
    \centering
    \includegraphics[width=1.0\textwidth]{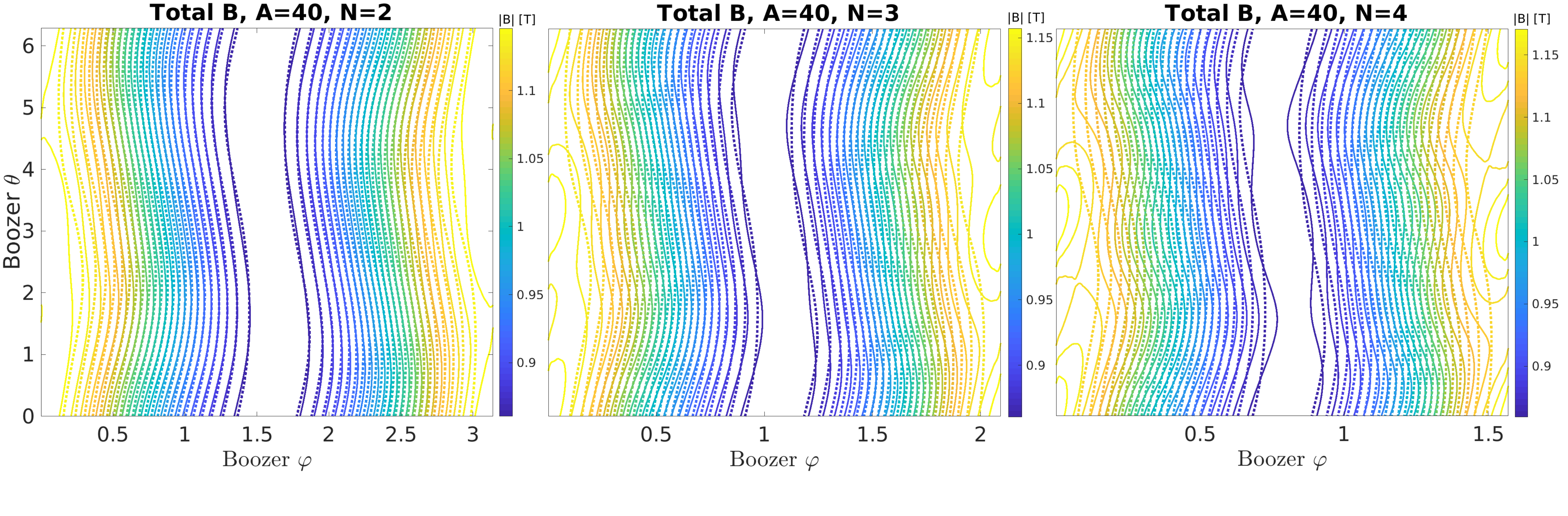}
    \caption{Contours of the magnetic field intensity on the plasma boundary, $B$, are shown for increasing values of $N$, the number of field periods, starting at $N=2$ on the left, $N=3$ center, $N=4$ right. Dotted lines show the contours obtained from the construction and solid lines those from the calculation with VMEC and BOOZ\_XFORM. Aspect ratio was set to $A=40$ for all cases.}
    \label{fig:B_comp_FieldPeriods_Pfefferle}
\end{figure}

The resulting boundary shape is used to find the magnetic field strength on the boundary using the MHD equilibrium code VMEC, and BOOZ\_XFORM. The result is compared with the magnetic field from the construction (eqn.~\ref{eq:magneticField}), and shown for A=40 in figure \ref{fig:B_comp_FieldPeriods_Pfefferle}, where it is clear that the approximation deteriorates with increasing number of field periods. 

\begin{figure}
    \centering
    \includegraphics[width=0.8\textwidth]{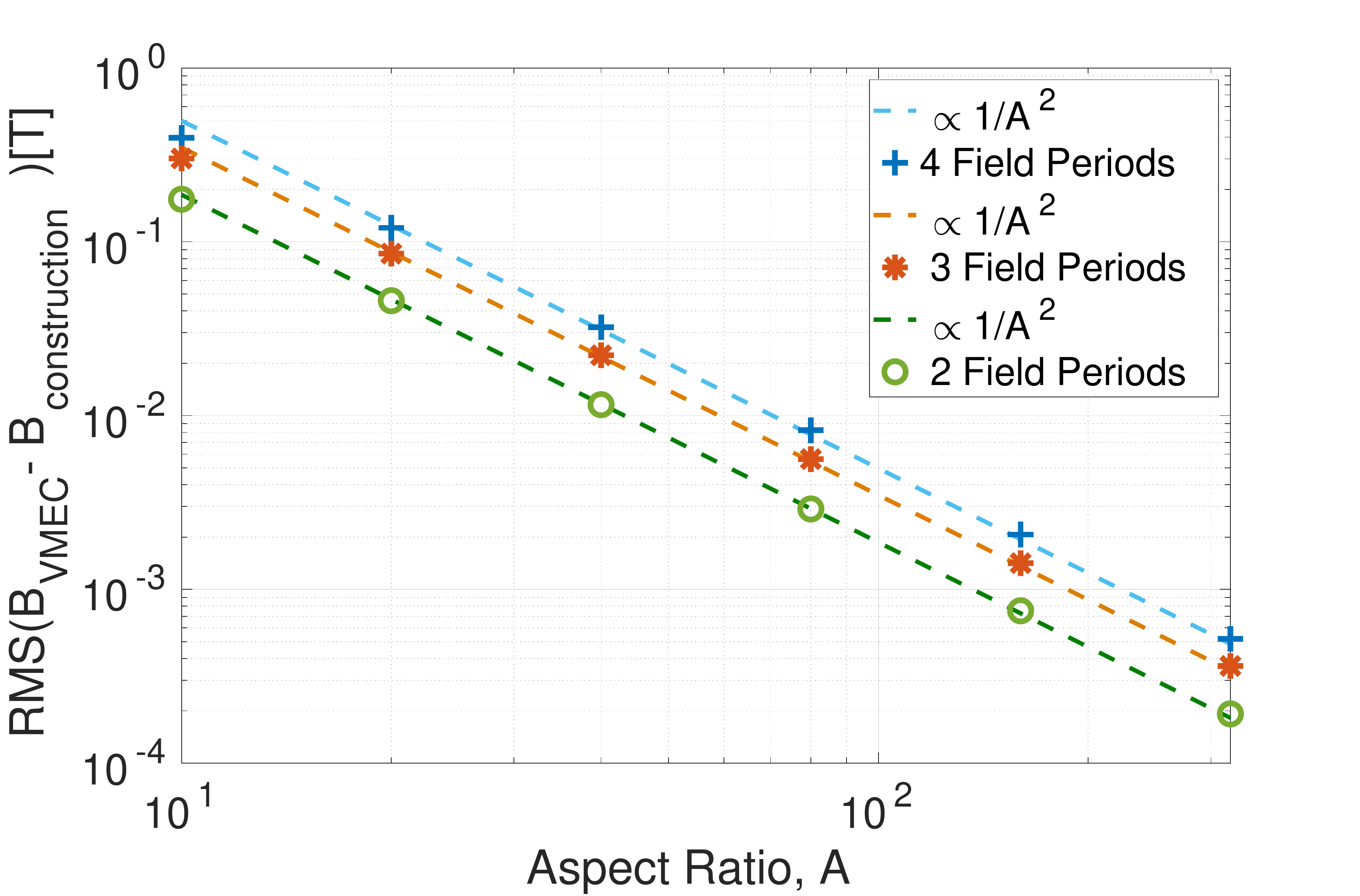}
    \caption{Root-mean-squared difference between the magnetic field from the construction and that calculated by VMEC for configurations with $N=2,3,4$ and different values of $A$. The dashed lines show the expected scaling $\propto 1/A^{2}$.}
    \label{fig:Brms_Pfefferle}
\end{figure}

\begin{figure}
    \centering
    \includegraphics[width=0.8\textwidth]{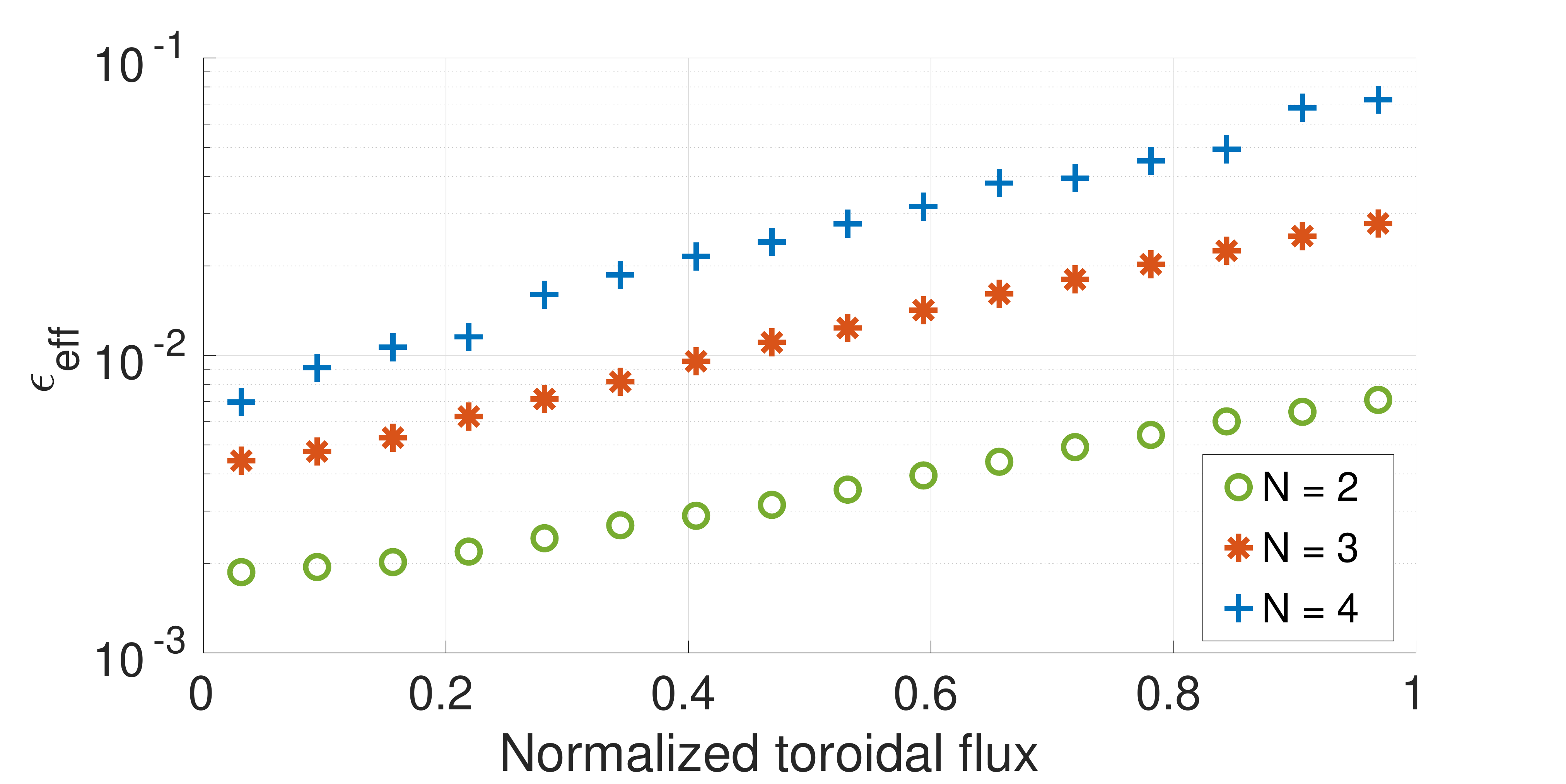}
    \caption{Effective ripple, $\epsilon_{\mathrm{eff}}$, profiles for configurations with $N=2,3,4$ field-periods.} 
    \label{fig:eps_eff_Pfefferle}
\end{figure}

In the same manner as in the previous section, we calculate the root-mean-square difference between the intensity of $B(\theta,\varphi)$ in the boundary as calculated by VMEC and as expected from the construction. This calculation is done for each configuration at different aspect ratios, and is shown in figure \ref{fig:Brms_Pfefferle}. The scaling for all cases is proportional to $1/A^2$, as expected, but the magnitude of the difference increases with the number of field periods, indicating a deterioration of the approximation with increasing $N$. The same behaviour is observed in the effective ripple, getting significantly worse for the case with 4 field periods and being optimal, below $\epsilon_{\mathrm{eff}}= 8\times 10^{-3}$, for two field periods, evident in figure \ref{fig:eps_eff_Pfefferle}. 

The only apparent significant differences between the  solutions constructed in this section are the number of field periods and the value of the torsion. Stellarator designs constructed through conventional optimisation have $N {=} 4,5,6$ and larger, including W7-X, a QI design (very approximately) with 5 field periods. This indicates there should be no fundamental obstacle to obtaining a good approximation to QI fields at larger values of $N$. Thus, with the hope to find such fields with the near-axis framework, we are further motivated to explore magnetic axes with low torsion. 

\section{A three-field-period, low-torsion-axis example}\label{3FP_example}

\begin{figure}
    \centering
    \includegraphics[width=0.8\textwidth]{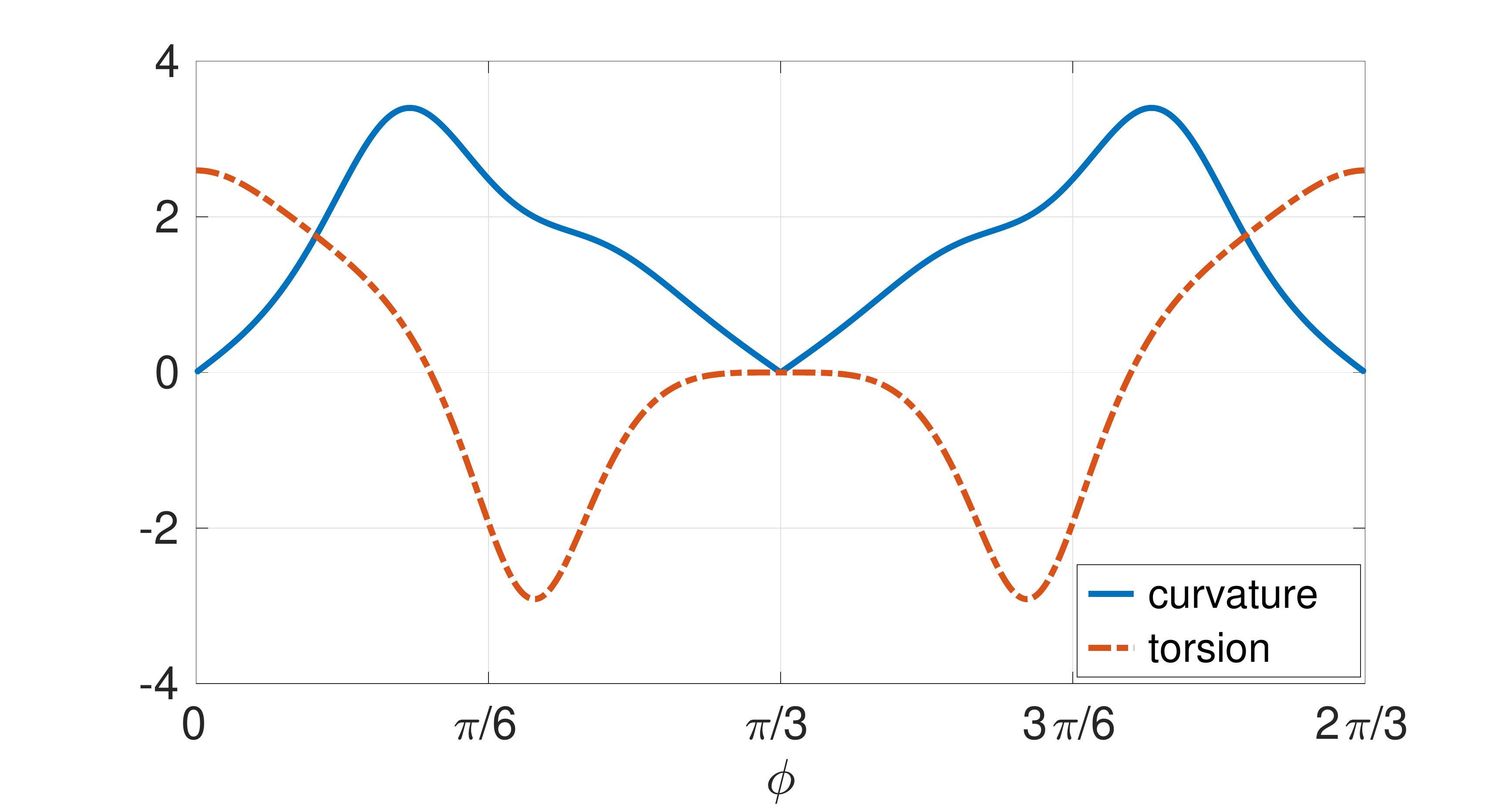}
    \caption{Curvature, $\kappa$, and torsion, $\tau$, with respect to the toroidal angle $\phi$ of the magnetic axis used in the construction of an $N=3$ equilibria. Curvature has zeros of first order at multiples of $\pi/3$, correponding to extrema of $B$. Torsion has a second order zeros at minima of B, $\varphi_{\mathrm{min}}^{i}$.}
    \label{fig:kappa_tau_3FP}
\end{figure}

Using the procedure described in section \ref{Axis_Shape}, we find closed curves with $N{=}3$ and zeros of first order in the curvature at extrema of $B(\varphi)$, and zeros of torsion at second order at $\varphi_{min}^{i}$. Then, motivated by the result of previous section, we choose one axis shape from this class that fulfils \begin{equation*}
    \mathrm{max_{\varphi}}(\tau) < \mathrm{max_{\varphi}}(\kappa)
\end{equation*} 
for all toroidal points. The Fourier coefficients describing this axis are
\begin{multline*}
        R = 1 + 9.07\times10^{-2} \cos{(3\phi)} - 2.06\times10^{-2} \cos{(6\phi)} 
        \\ - 1.11\times10^{-2} \cos{(9\phi)} - 1.64\times10^{-3} \cos{(12\phi)},
\end{multline*}
\begin{equation*}
    z = 0.36 \sin{(3\phi)} + 2.0\times10^{-2}\sin{(6\phi)}+ 1.0\times10^{-2}\sin{(9\phi)}.
\end{equation*}
Its curvature and torsion are shown as functions of the toroidal angle $\phi$ in figure \ref{fig:kappa_tau_3FP}, and $m=0$.

\begin{figure}
    \centering
    \includegraphics[width=1.0\textwidth]{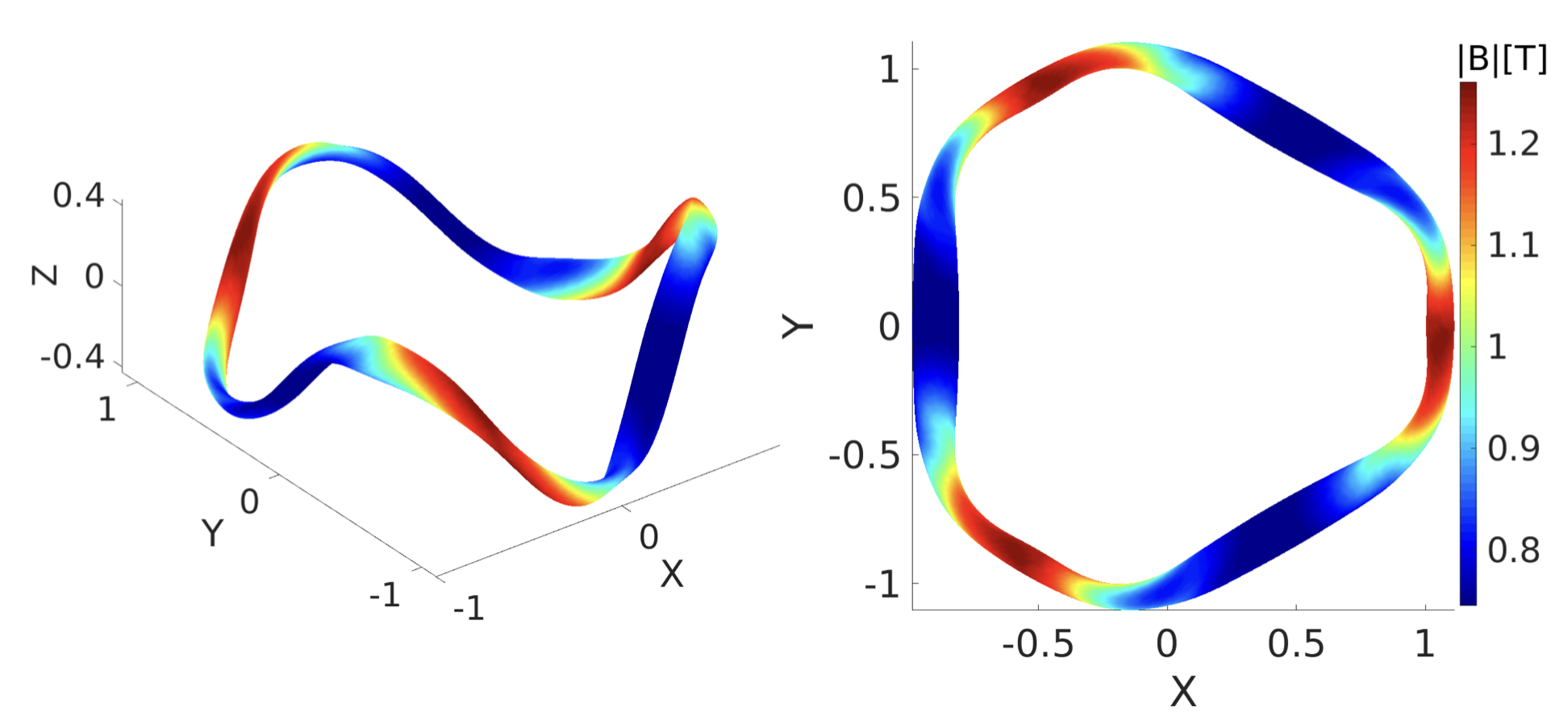}
    \caption{Intensity of the magnetic field on the plasma boundary for a 3 field period, A=20 configuration. Side view (left) and top view (right) are shown.}
    \label{fig:Boundary_3FP}
\end{figure}

\begin{figure}
\centering
  \includegraphics[width=0.5\textwidth]{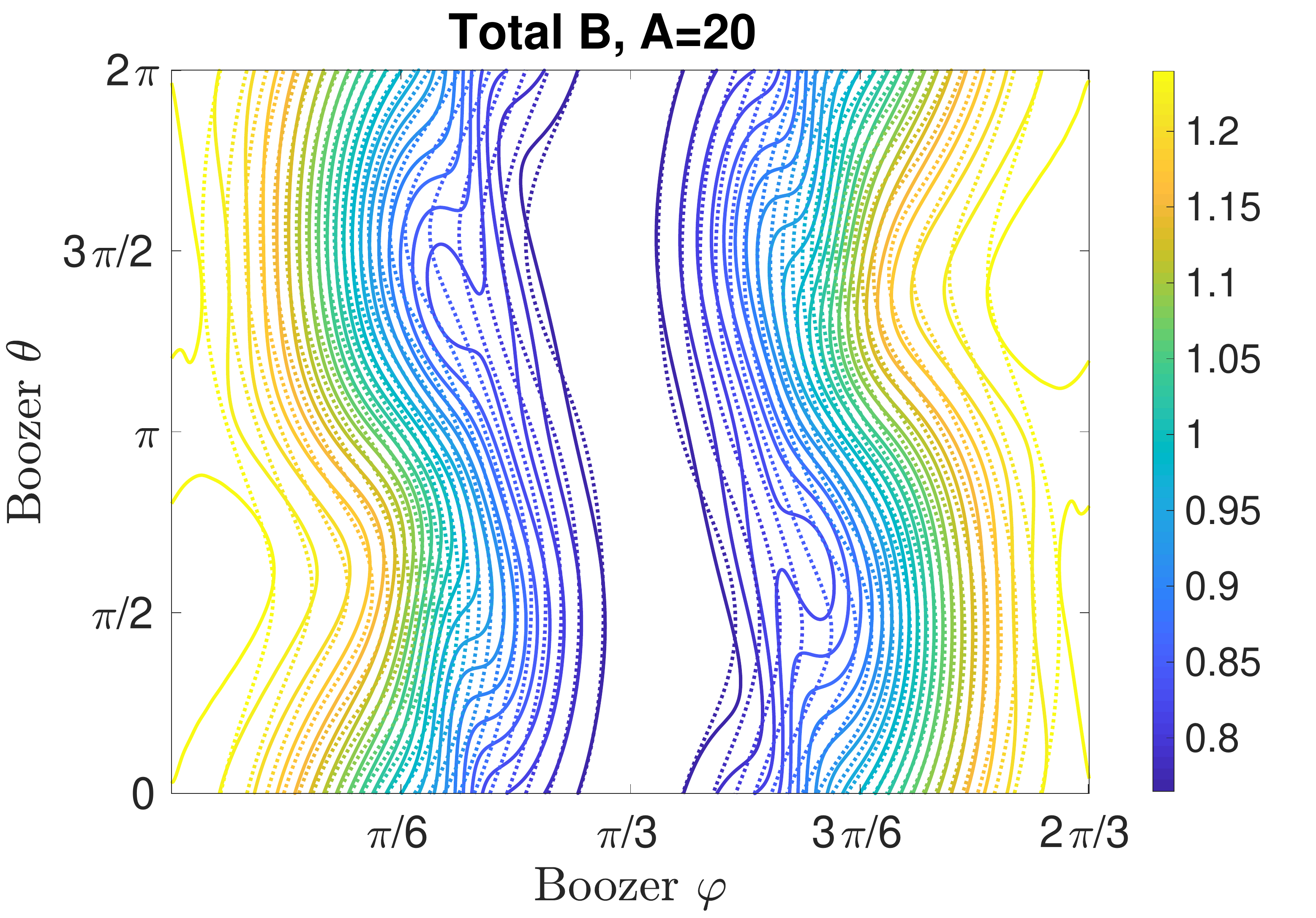}
\caption{ Contours of the magnetic field intensity on the boundary, for $N=3$, $A=20$, as calculated by VMEC (solid lines) and from the near-axis construction(dotted lines).}
\label{fig:Btot_contours_3FP}
\end{figure}

Using this axis, an $N=3$, QI boundary was constructed for $A=20$, with a magnetic field on-axis given by
\begin{equation*}
    B_{0} = 1 + 0.25\cos{(3\varphi)},
\end{equation*}
$d(\varphi) = 1.03\kappa^{s} $ and the parameter $k$ entering eqn.(\ref{eq:alpha_final}) set to $2$. The resulting plasma boundary and the magnetic field strength on it as found by VMEC are shown in figure \ref{fig:Boundary_3FP}, where we can observe straight sections around $\varphi_{\mathrm{min}}$, as a consequence of the axis choice. Figure \ref{fig:Btot_contours_3FP} shows the contours of $|B|$ as calculated by VMEC and from the near-axis construction in Boozer coordinates. At this aspect ratio most of the contours still close poloidally, as necessary for quasi-isodynamicity. The magnetic field contours are less straight around the point of maximum $B_{0}$, which is expected since this is the region where $\alpha(\varphi)$ deviates from a perfectly omnigenous form. 

\begin{figure}
\centering
  \includegraphics[width=1.0\textwidth]{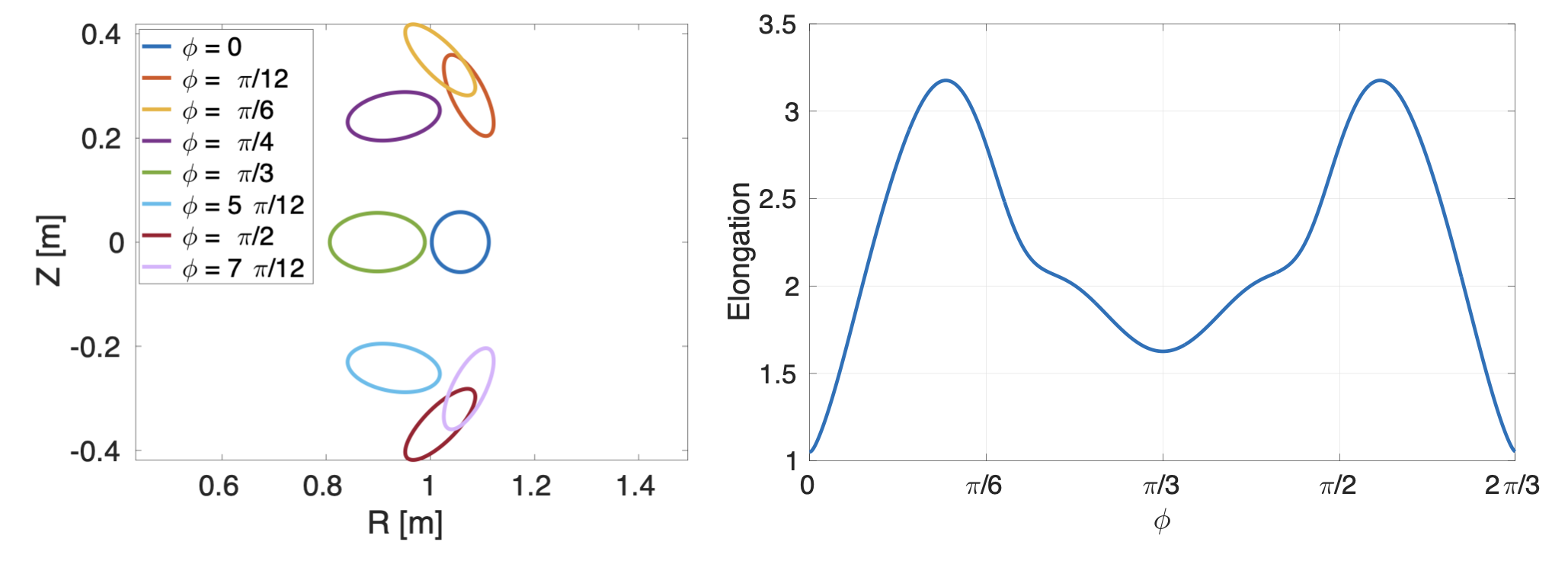}
\caption{ (left) Cross sections of the plasma boundary for different values of the poloidal angle $\phi$, and (right) elongation with respect to the cylindrical toroidal angle $\phi$. }
\label{fig:elong_3FP_contours}
\end{figure}

 Zero torsion around the bottom of the magnetic well was chosen, together with $d(\varphi)$, as an attempt to reduce the elongation of the plasma boundary, as described in section \ref{Axis_Shape}. In figure \ref{fig:elong_3FP_contours} we see the cross-sections of the plasma boundary for different values of toroidal angle on the left and the evolution of elongation with the toroidal angle. The maximum elongation for this configuration is $e_\mathrm{max}{=}3.2$, and includes regions where the elongation is nearly $e=1$, the case of a circular cross-section. We also observe that the elongation remains low around the region where torsion vanishes. These facts seem to validate our approach for controlling elongation, and demonstrate that high elongation is not, as might be feared from previous results, a necessary sacrifice for achieving good omnigenity. 

\begin{figure}
\centering
  \includegraphics[width=0.49\textwidth]{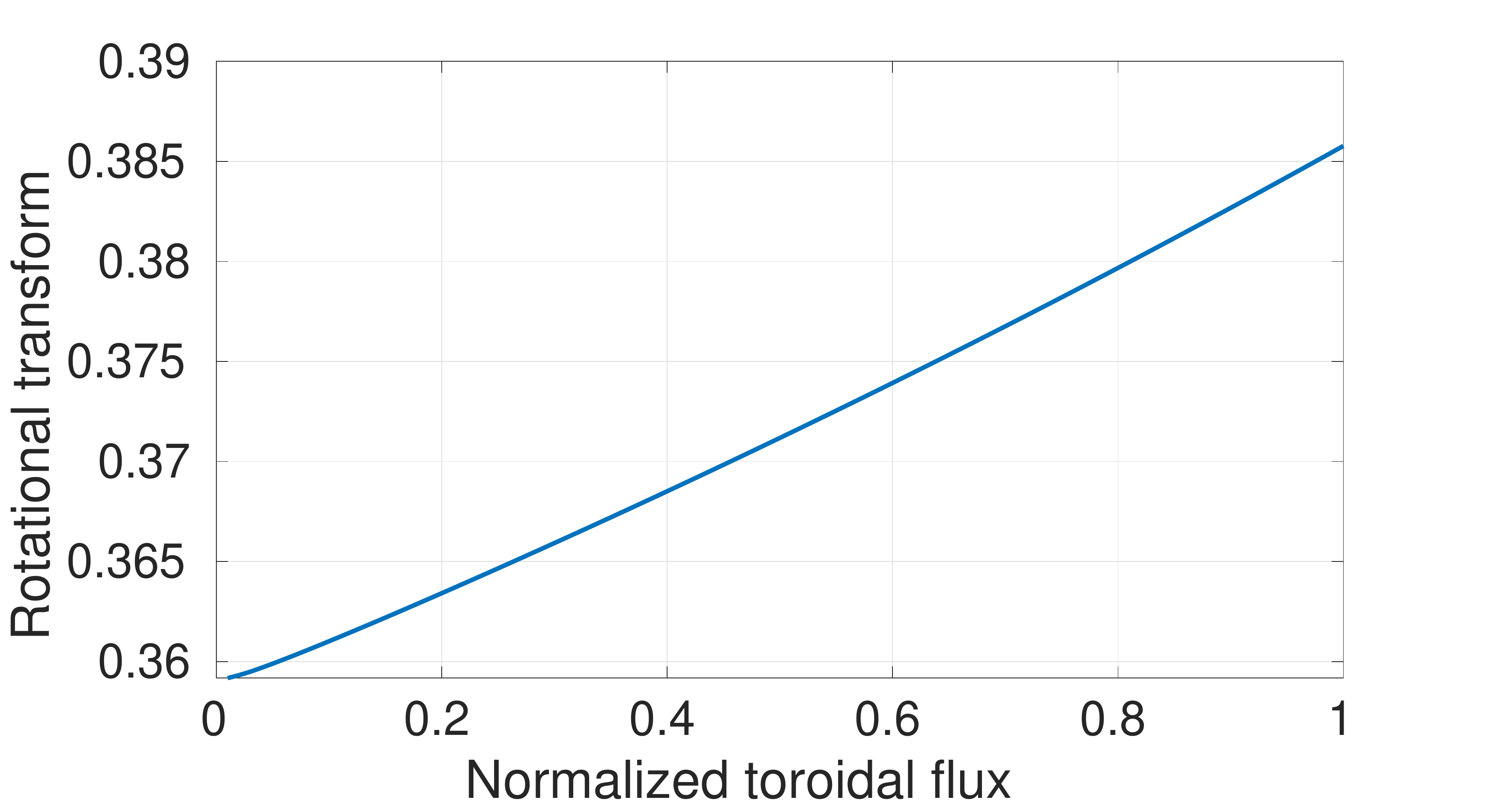}
  \includegraphics[width=0.49\textwidth]{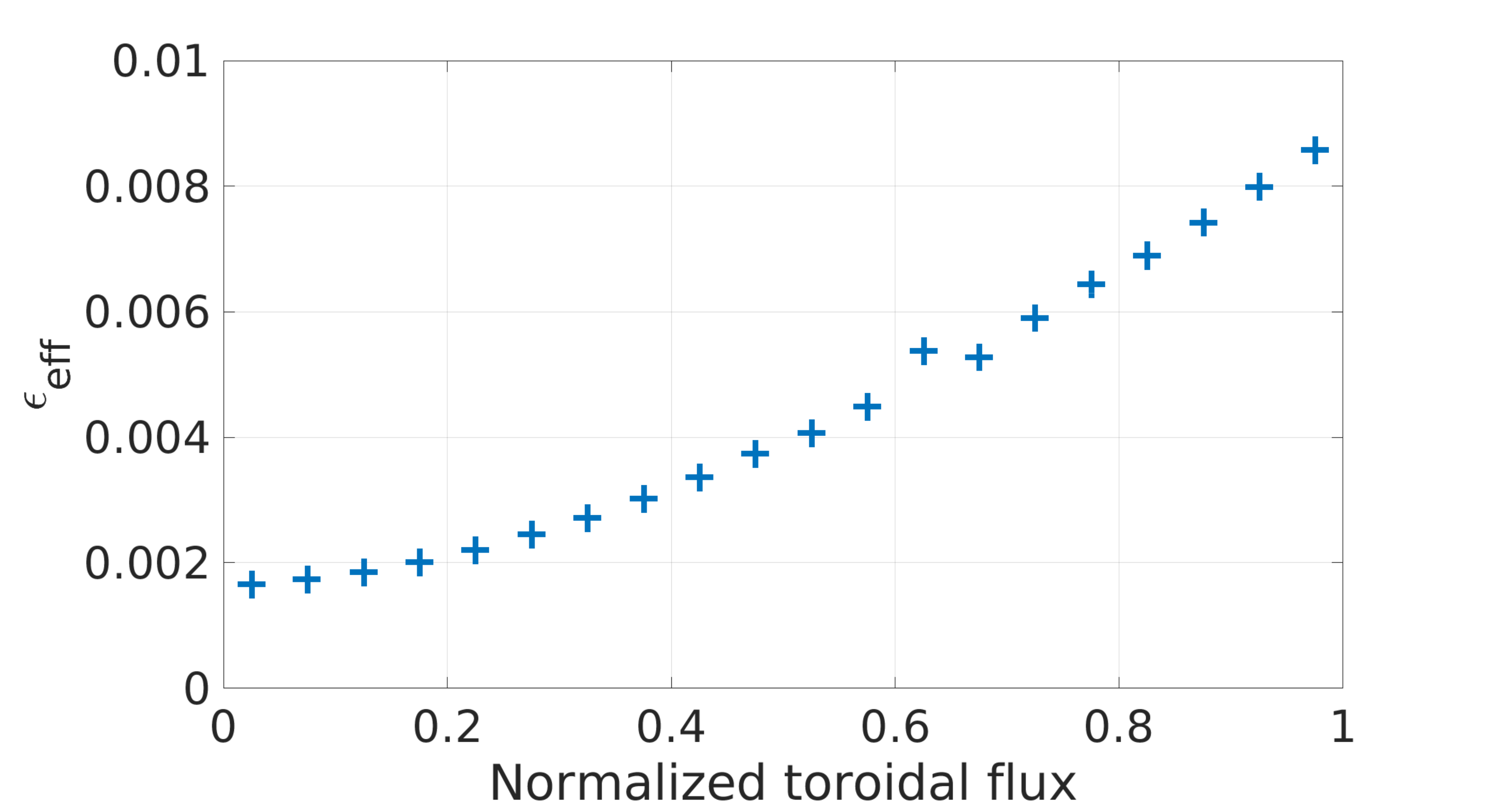}
\caption{ (left) Rotational transform, $\iotaslash$, and  (right) Effective ripple $\epsilon_{\mathrm{eff}}$} for an $N=3$, $A=20$ configuration.
\label{fig:iota_3FP_eps_eff}
\end{figure}

 The rotational transform profile obtained with VMEC is shown in figure (\ref{fig:iota_3FP_eps_eff}) and coincides with the value calculated numerically from Eq.~(\ref{eq:sigma}), $\iotaslash = 0.375$. The effective ripple, $\epsilon_{\mathrm{eff}}$,  was found to be lower than 1\% throughout the plasma volume, another indication of the closeness of the solution to omnigenity. The value at different distances from the axis is shown in Fig.~(\ref{fig:iota_3FP_eps_eff})(left).

\section{Conclusion}

We have described in detail the near-axis-expansion method to construct QI, stellarator symmetric, single-magnetic-well equilibria with $N\ge 1$ field periods, valid at first order in the distance from the magnetic axis. A new way of achieving better continuity and smoothness of these configurations, as compared with the previous method of \cite{plunk2019direct}, is introduced and used to construct equilibria with $N>1$. 

The problem of finding axis shapes compatible with the near-axis expansion is discussed, in particular the order of zeros in axis curvature, and the naturally arising increase of torsion for increasing number of field periods, which we argue is an underlying reason for the  deterioration of the approximation at finite aspect ratio. A method to systematically describe and construct closed curves with zeros of curvature and torsion, at different orders, at toroidal locations of extrema of the magnetic field strength is also presented.  

 We demonstrate the validity of the near-axis-expansion method, with a two-field-period example, by showing that the difference between the magnetic field as calculated by the NAE and that obtained using the equilibrium code VMEC falls with increasing aspect ratio, and scales as $1/A^2$, as expected from a first-order expansion. We describe the construction of this two-field-period configuration, and find that it has good confinement, as shown by $\epsilon_{\mathrm{eff}}<2\%$ throughout the plasma volume at aspect ratio $A{=}10$.
 
We also construct a family of solutions, for $N=2,3,4$, having axes with constant torsion, and very similar per-field-period initial parameters. The approximate omnigenity of these solutions deteriorates if the the number of field periods is increased. This example shows the importance of reducing the maximum torsion of the axis to achieve equilibria close to omnigenity at low aspect ratio. 

In the last section we demonstrate how the choice of an axis with zero torsion around the point of minimum magnetic field strength and constrained maximum torsion enables us to find a three-field-period configuration with low elongation and small neoclassical transport. The effective ripple remains at under $1\%$ for an aspect ratio $A=20$, and the maximum elongation $e_{max}=3.2$, demonstrating that low elongation is achievable in QI stellarators.

These configurations demonstrate that the near-axis-expansion method can be used to construct magnetic equilibria with multiple field periods that maintain good confinement properties at low aspect ratios. We emphasize that these examples were all obtained without the need of numerically costly optimisation procedures.

Given the importance of the axis shape in the quality of the equilibrium, a natural next step is to reintroduce an element of optimisation, as done for the one-field-period example of \cite{Jorge2022}, but appropriately restricting the search space. Specifically, we can define an optimisation space according to classes of axis curves satisfying prescribed conditions on the torsion and curvature, and search this space for configurations with attractive properties. Another interesting analysis, which can be performed thanks to the speedy calculation of solutions enabled by the near-axis expansion, is to systematically  and exhaustively map the space of QI solutions and its dependence on the input functions used for the construction. Such exploration might allow physical insight to be gained into the structure of the solution space and help explain certain difficulties associated with traditional optimisation techniques.    

The only shaping of the plasma boundary that enters at first order in the NAE is the elongation of the elliptical cross-sections. Using solutions generated at higher order, together with traditional optimisation procedures also seems a promising way for obtaining configurations with better confinement properties and stronger shaping.  

\section*{Acknowledgments}

The authors would like to thank Matt Landreman for providing the numerical code, described in \cite{plunk2019direct}, which was adapted for the present study. We are also grateful to Michael Drevlak for providing the code used for calculating the effective ripple and for fruitful discussions. This work was partly supported by a grant from the Simons Foundation (560651, KCM).

\section{Appendix I. Smoother $\alpha(\varphi)$}
The behaviour of $\alpha(\varphi)$ around $\varphi_{\mathrm{max}}$, where omnigenity is broken in a controlled way, can have a detrimental impact on the smoothness of the QI solutions constructed using the near-axis expansion. 

In order to avoid this problems, a form of $\alpha$ with continuous derivatives up to second order is proposed
\begin{equation}\label{eq:alpha_app}
    \alpha_{\mathrm{II}}(\varphi) = \iotaslash (\varphi-\varphi_{min}^{i})+\pi (2 m i + \tfrac{1}{2}) + a(\varphi-\varphi_{min}^{i})^{2k+1} + b(\varphi-\varphi_{min}^{i})^{2p+1} 
\end{equation} 
To find the coefficients $a$ and $b$, we check for continuity at $\varphi_{\mathrm{max}}^{i}$

\begin{equation*}
    \lim_{\varphi\to\varphi_{max}^{(i+1)-}} \alpha_{\mathrm{II}}(\varphi)  = \lim_{\varphi\to\varphi_{max}^{(i+1)+}} \alpha_{\mathrm{II}}(\varphi),
\end{equation*}
which is equivalent to
\begin{multline*}
    \iotaslash \left( - \pi / N\right) + a\left(-\pi / N\right)^{2k+1}  + b\left(-\pi / N\right)^{2p+1}+ 2\pi m \\= \iotaslash \left(\pi / N\right) + a\left(\pi / N\right)^{2k+1} + b\left(\pi / N\right)^{2p+1},  
\end{multline*}
and grouping terms we obtain
\begin{equation}\label{eq:a_b_rel}
 \pi \left(m-\frac{\iotaslash}{N}\right) \\= a\left(\pi / N\right)^{2k+1} + b\left(\pi / N\right)^{2p+1}.  
\end{equation}
From expression (\ref{eq:alpha_app}), we note that all even derivatives of $\alpha_{\mathrm{II}}$ with respect to $\varphi$ have odd powers of $(\varphi-\varphi_{min}^{i})$, so $d^{2n}\alpha_{\mathrm{II}}/d\varphi^{2n}$ is opposite in sign, for $n\geq1$, at $\varphi^{i}_{\mathrm{max}}$ when approaching from the left and from the right, hence continuity requires all even derivatives to be zero at this points. We now impose the following condition on the second derivative
\begin{equation*}
    \frac{d^{2}\alpha_{\mathrm{II}}(\varphi_{\mathrm{max}})}{d\varphi^{2}} = 0,
\end{equation*}
\begin{equation*}
    a(2k+1)(2k)(\pi/N)^{2k-1} + b(2p+1)(2p)(\pi/N)^{2p-1}= 0.
\end{equation*}
and solve for $b$,
\begin{equation*}
    b = -a \nu (\pi/N)^{2k-2p}= 0,
\end{equation*}
where $\nu$ is given by
\begin{equation*}
    \nu = \frac{(2k+1)(2k)}{(2p+1)(2p)}. 
\end{equation*}
Next, we substitute this expression for $b$ in equation (\ref{eq:a_b_rel})
\begin{equation*}
    \pi \left(m-\frac{\iotaslash}{N}\right) \\= a \left(  1 - \nu \right) \left(\frac{\pi}{N}\right)^{2k+1},
\end{equation*}
from where we find an expression for $a$
\begin{equation}\label{eq:a_app}
    a = \pi \left(m-\frac{\iotaslash}{N}\right)\left( \frac{1}{1-\nu}\right) \left(\frac{N}{\pi}\right)^{2k+1}, 
\end{equation}
and following the same process we obtain an expression for $b$
\begin{equation}\label{eq:b_app}
    b = -\pi \left(m-\frac{\iotaslash}{N}\right)\left( \frac{\nu}{1-\nu}\right) \left(\frac{N}{\pi}\right)^{2p+1}. 
\end{equation}
For the case of odd derivatives of $\alpha_{\mathrm{II}}$, we obtain even powers of $(\varphi-\varphi_{min}^{i})$, so all odd derivatives are are automatically continuous at $\varphi_{\mathrm{max}}$ due to symmetry of the magnetic wells. As a consequence, expression (\ref{eq:alpha_app}) with $a$ and $b$ given by Eq.~(\ref{eq:a_app}) and Eq.~(\ref{eq:b_app}) is smooth and continuous up to third order derivatives. 

\section{Appendix II. Axis Shapes}
We use the Fourier axis representation as described in section \ref{Axis_Shape}. 
A local form is also needed to establish conditions on the derivatives of these functions about a point of stellarator symmetry (also coinciding with an extrema of $B_0(\phi)$). This local form can be then used to generate a linear system of equations on the Fourier coefficients.  These points are given at $\phi = n \pi/N$ for arbitrary integer $n$, but without loss of generality we take the point to be $\phi =0$ (and perform any necessary shifts later):

\begin{eqnarray}
    R(\phi) = \sum_{i=0} \frac{R_{2i}}{(2i)!} \phi^{2i},\label{R-series}\\
    z(\phi) = \sum_{i=0} \frac{z_{2i+1}}{(2i+1)!} \phi^{2i+1}.\label{z-series}
\end{eqnarray}
From the stellarator-symmetric forms of $R$ and $z$, one significant fact should be noted -- the only stellarator-symmetric planar curves are those with $z=0$ for all $\phi$ (excluding the trivial `tilted' one with $N=1$).  As we will see, stellarator-symmetric axis curves that are consistent with omnigenity require $z\neq 0$, and therefore must be non-planar, and possess finite torsion.  (As experience shows, attempts to tune the axis shape for low torsion in one region, result in large torsion elsewhere.)

The curvature and torsion of general parameterization of a curve are given by
\begin{eqnarray}
    \kappa = \frac{|{\bf x}^\prime \times {\bf x}^{\prime\prime}|}{|{\bf x}^\prime|^3},\\
    \tau = \frac{({\bf x}^\prime\times{\bf x}^{\prime\prime})\cdot{\bf x}^{\prime\prime\prime}}{|{\bf x}^\prime\times{\bf x}^{\prime\prime}|^2},
\end{eqnarray}
where primes denote differentiation with respect to $\phi$.  Noting $d\hat{\bf R}/d\phi = \hat{\bf \phi}$ and $d\hat{\bf \phi}/d\phi = -\hat{\bf R}$, it is straightforward to compute these derivatives.  Further substituting the expansions for $R$ and $z$, Eqs.~\ref{R-series}-\ref{z-series}, the contributions to each derivative can be collected by their order in $\phi$, the following (possibly useful) properties can be confirmed:

\begin{eqnarray}
    \hat{\bf R}\cdot\left(\frac{d^n{\bf x}}{d\phi^n}\right)_m = 0\text{, for odd } n + m,\\
    \hat{\bf z}\cdot\left(\frac{d^n{\bf x}}{d\phi^n}\right)_m = \hat{\bf \phi}\cdot\left(\frac{d^n{\bf x}}{d\phi^n}\right)_m = 0\text{, for even } n + m.
\end{eqnarray}
where the subscript $m$ denotes the order in $\phi$.

We will classify the zeros in curvature and torsion by the order of the first non-zero term in the power series, for example 

\begin{equation}
\kappa = \kappa_m \phi^m + \kappa_{m+1} \phi^{m+1} +\dots,
\end{equation}
where it is assumed that $\phi > 0$; this is necessary to fix the sign of the coefficients noting that $\kappa > 0$ by convention.  Likewise, the first non-zero term in the power series expansion of $\tau$ determines its order:

\begin{equation}
\tau = \tau_n \phi^n + \kappa_{n+1} \phi^{n+1} +\dots,
\end{equation}

We can denote the two constraints by the pair $(m, n)$ corresponding to the order of the curvature and torsion zeros, respectively.

\subsection{Conditions on curvature}

Assuming that ${\bf x}^\prime$ is itself non-zero, the conditions on zeros in curvature are found by requiring ${\bf x}^\prime \times {\bf x}^{\prime\prime} = 0$ at each order in $\phi$.  At zeroth order, ${\bf x}^{\prime\prime}$ has its only non-zero component in the $\hat{\bf R}$ direction, and the condition ${\bf x}^\prime \times {\bf x}^{\prime\prime} = 0$ is satisfied by ${\bf x}^{\prime\prime}\cdot\hat{\bf R} = 0$.  The curvature can be made zero to higher order by considering higher-order contributions to ${\bf x}^{\prime\prime}$.  At odd orders, these are contained in the $\hat{\bf \phi}-\hat{\bf z}$ plane and must be made parallel to the zeroth order contribution from ${\bf x}^\prime$, while at even orders, the even order contribution to ${\bf x}^{\prime\prime}$ must simply be zero. Thus, conditions at arbitrary order can be obtained, and a few are listed below\\

\begin{table}
\begin{center}
\begin{tabular}{c c}
\hline
 Order & Constraint \\
\hline
 $0$ & $R_2=R_0$ \\
 $1$ & $z_3=2 z_1$ \\
 $2$ & $R_4=5 R_0$ \\
 $3$ & $z_5=16 z_1$ \\
 $4$ & $R_6=61 R_0$ \\
\hline
\end{tabular}
\caption{Conditions for zero curvature.}
\label{curvature-table}
\end{center}
\end{table}

As already noted, only odd-order zeros in curvature are consistent with omnigenity in the near-axis expansion.  Thus, we apply these conditions only up to and including some even order.  Note that planar curves (for which $z = 0$) are inconsistent with odd orders of zero in curvature, which means non-zero torsion is required for the class of configurations being considered here.

\subsection{Conditions on torsion}

We find curves with zero torsion to some order of accuracy in the local expansion about points of stellarator symmetry.  It is assumed that the curvature is zero to some order at these points.  This implies that curves of zero torsion, which are approximately planar curves, fall into one of two classes: (1) curves lying within the plane perpendicular to $\hat{\bf x} = \hat{\bf R}(0)$ and (2) curves lying within the plane perpendicular to a constant unit vector $\hat{\bf n}(t) = \cos(t)\hat{\bf z} + \sin(t)\hat{\bf y}$, where $\hat{\bf y} = \hat{\bf \phi}(0)$, and $t$ is arbitrary.  

The two classes arise in the expansion itself: let us inspect the first non-zero contribution to the numerator and denominator of the expression for the torsion (assuming a first order zero of curvature, i.e. $R_2 = R_0$):

\begin{eqnarray}
     ({\bf x}^\prime\times{\bf x}^{\prime\prime})\cdot{\bf x}^{\prime\prime\prime} \approx \frac{1}{2} R_0 \left(5 R_0-R_4\right) \left(2 z_1-z_3\right)\phi^2,\\
     |{\bf x}^\prime\times{\bf x}^{\prime\prime}|^2 \approx R_0^2 \left(2 z_1 - z_3\right)^2\phi^2,
\end{eqnarray}
which assuming $z_3 \neq 2 z_1$ yields a result for the torsion at $\phi = 0$:

\begin{equation}
    \lim_{\phi \rightarrow 0}\tau = \frac{5 R_0-R_4}{4 R_0 z_1-2 R_0 z_3}
\end{equation}

Thus, if the curvature is zero to first order, but not second order, we obtain a condition on the torsion being zero, $5 R_0 = R_4$.  The second class of curves with zero torsion is obtained by repeating this calculation assuming $z_3 = 2 z_1$ from the outset, but this is precisely the condition that curvature is zero to second order; it is also the condition that the curve lies within the described plane, and it can be derived independently by imposing the condition $\hat{\bf n}\cdot{\bf x}^{\prime} = 0$ to first order in $\phi$.  Even-order zeros, however, are not consistent with near-axis QI configurations.

Below we calculate the conditions for the torsion, order-by-order, assuming a number of conditions are also satisfied related to curvature.  The relevant cases included in Table \ref{curvature-table} are first, third and fifth order zeros of curvature.  The torsion constraints are given in the tables below.

\begin{table}
\begin{center}
\begin{tabular}{c c}
\hline
 Order & Constraint \\
\hline
 $0$ & $R_4=5 R_0$ \\
 $2$ & $R_6=61 R_0$ \\
 $4$ & $R_8=1385 R_0$ \\
 $6$ & $R_{10}=50521 R_0$ \\
\hline
\end{tabular}
\caption{Conditions for zero torsion, assuming first order zero in curvature.}
\label{torsion-table-1}
\end{center}
\end{table}

\begin{table}
\begin{center}
\begin{tabular}{c c}
\hline
 Order & Constraint \\
\hline
 $0$ & $R_6=61 R_0$ \\
 $2$ & $R_8=1385 R_0$ \\
 $4$ & $R_{10}=50521 R_0$ \\
\hline
\end{tabular}
\caption{Conditions for zero torsion, assuming third order zero in curvature.}
\label{torsion-table-3}
\end{center}
\end{table}

\begin{table}
\begin{center}
\begin{tabular}{c c}
\hline
 Order & Constraint \\
\hline
 $0$ & $R_8=1385 R_0$ \\
 $2$ & $R_{10}=50521 R_0$ \\
 $4$ & $z_7=272 z_1$ \\
\hline
\end{tabular}
\caption{Conditions for zero torsion, assuming third order zero in curvature.}
\label{torsion-table-5}
\end{center}
\end{table}

\subsection{Truncated Fourier representations of axis curves}

The tabulated constraints on the derivatives of the axis components can be applied to a truncated Fourier representation.  Eqs.~\ref{R-Fourier}-\ref{z-Fourier} are simply substituted into the constraint equations with $\phi$ set to a location of stellarator symmetry (for instance at $\phi = 0, \pi/N$ in the first period).  This results in a set of linear conditions on the Fourier coefficients that can be solved numerically, or by computer algebra.  

As a very simple example, one may consider a symmetric class of curves, as in \cite{plunk2019direct} and described in section \ref{sec:Axis_Construction}.

Curve classes can of course also be defined without the above symmetry (retaining odd harmonics), requiring derivative constraints to be applied separately at $\phi = 0$ and $\phi = \pi/N$.  For example, consider

\begin{eqnarray}
     R = 1 + R_c(1) \cos(N \phi) + R_c(2) \cos(2 N \phi) + R_c(3) \cos(3 N \phi),\\
     z = z_s(1) \sin(N \phi) + z_s(2) \sin(2 N \phi) + z_s(3) \sin(3 N \phi).
\end{eqnarray}
Then, applying $R_2 = R_0$ at both $\phi = 0$ and $\phi = \pi/N$, gives

\begin{eqnarray}
R_c(1) = -R_c(3) \frac{9 N^2 + 1}{N^2+1},\\
R_c(2) = -\frac{1}{4 N^2+1}
\end{eqnarray}
These are just the simplest cases, with only a zeroth order condition for curvature being used.  Note that different sets of derivative constraints can be applied to these two locations, to get curves with different orders of zeros in torsion and curvature at the locations of maximum and minimum magnetic field strength.  The orders of the zeros, together with the set of Fourier coefficients, define a space that can be used for further optimisation.

\bibliographystyle{jpp}
\bibliography{omnigenity-near-axis}

\end{document}